\documentclass[letterpaper,12pt]{JHEP}
\usepackage{epsfig}
\newcommand{\embedsp}[1]{{\tilde{#1}}}
\newcommand{\embedvec}[1]{{\tilde{#1}}}
\newcommand{\cmplxconj}[1]{{\overline{#1}}}
\newcommand{\dualrep}[1]{{ {{#1}}}}

\newcommand{\beq}{\begin{equation}}
\newcommand{\eeq}{\end{equation}}
\newcommand{\bea}{\begin{eqnarray}}
\newcommand{\eea}{\end{eqnarray}}
\newcommand{\conjrep}[1]{{\tilde{#1}}}

\title{Riemannian Gauge Theory and Charge Quantization}
\author{ Mario Serna$^a$ and Kevin Cahill$^b$ \\
 \small $^a$\ Space Vehicles Directorate, Air Force Research Labs\\
 \small 3550 Aberdeen Ave, Kirtland, New Mexico 87117-5776, USA\\
 \small E-mail: mariojr@alum.mit.edu\\
 \small $^b$\ Department of Physics and Astronomy, University of New
Mexico\\
 \small Albuquerque, New Mexico 87131-1156, USA\\
 \small E-mail: cahill@unm.edu}

\abstract{
In a traditional gauge theory,
the matter fields $\phi^a$ and
the gauge fields $A^c_\mu$ are fundamental objects
of the theory.
The traditional gauge field
is similar to the connection coefficient in the
Riemannian geometry covariant derivative, and the field-strength
tensor is similar to the curvature tensor.
In contrast, the connection in Riemannian geometry is derived
from the metric or an embedding space.
Guided by the physical principal of increasing symmetry among
the four forces, we propose a different
construction.  Instead of defining the transformation
properties of a fundamental gauge field,
we derive the gauge theory from an embedding of a gauge fiber
\(F={\bf R}^n\) or \(F={\bf C}^n\)  into a trivial, embedding
vector bundle \(\embedsp F={\bf R}^N\) or \(\embedsp F={\bf C}^N\) where \(N>n\).
Our new action is symmetric between the gauge theory and the Riemannian geometry.
By expressing gauge-covariant fields in terms of the
orthonormal gauge basis vectors, we recover a traditional, $SO(n)$ or $U(n)$
gauge theory.
In contrast, the new theory has all matter fields on a particular
fiber couple with the same coupling constant.
Even the matter fields on a ${\bf C}^1$ fiber, which have a $U(1)$ symmetry group,
couple with the same charge of $\pm q$.
The physical origin of this unique coupling constant
is a generalization of the general relativity equivalence principle.
Because our action is independent of the choice of basis,
its natural invariance group is $GL(n,{\bf R})$ or
$GL(n,{\bf C})$\@.
Last, the new action also requires a small correction to the
general-relativity action proportional to
the square of the curvature tensor.}


\begin{document}
% \maketitle

\section{Traditional Gauge Theory}
\label{SecGaugeTheoryReview}

In a traditional gauge theory~\cite[p.~7]{Weinberg:1996},
the matter fields $\phi^a$ and
the gauge fields $A^c_\mu$ are fundamental objects
of the theory with
$(A_\mu)^a_{\ b}(x)
= (T^c)^a_{\ b} A_\mu^c(x)$,
where the $T^c$ are the generators of the gauge group \(G\).
The gauge theory is defined
by the properties given to these fundamental fields.
One defines a continuous local symmetry group $G$\,
and a transformation rule $\phi'^a(x) = g^a_{\ b}(x)\, \phi^b(x)$
for every multiplet $\phi^a(x)$ of matter fields,
where \(g(x)\) is some representation of \(G\).
One maintains
the gauge invariance of the matter-field
derivative by introducing a gauge
field $A_\mu$ and a gauge-field transformation rule
 \begin{equation}
 (A'_\mu)^a_{\ b} = g^a_{\ c} \, (A_\mu)^c_{\ d} \,
{(g^{-1})}^d_{\ b} + (i/q) \, g^a_{\ c} \,\partial_\mu \, {(g^{-1})}^c_{\ b}
 \label{EqATransfRule}
 \end{equation}
that cancels the extra term appearing in the derivative
$\partial_\mu (g\,\phi)$\@.
These properties
are collected and clearly defined in terms of
principal bundles \cite{96Nakahara}.
In the action,  one includes every Lorentz-invariant,
gauge-invariant, renormalizable interaction.
\par
The traditional gauge theory with a fundamental gauge field
has many parallels with
Riemannian geometry~\cite{56Utiyama,Eguchi:1980jx,Gambini:1981yz}.
The covariant derivatives of a matter field $\phi^a$ appear
together with the gauge fields, $D_\mu \phi^a = (\delta^a_{b} \partial_\mu - i q
(A_\mu)^a_{\ b} ) \phi^b$, in a structure that resembles the covariant derivatives
of a vector $V^\nu$ in Riemannian geometry
$\nabla_\mu V^\nu = (\delta^\sigma_\mu \partial_\sigma + \Gamma^\mu_{\nu \sigma}) V^\sigma$.
The field-strength tensor $F_{\mu \nu}$,
which is constructed
to maintain the gauge invariance of
${\rm Tr}(F_{\mu \nu} F^{\mu\nu})$,
is the commutator of two covariant derivatives
 \begin{equation}
  -i\,{F_{\mu \nu}}^a_{\ b} \phi^b = ([ D_\mu , D_\nu ] \phi)^a
  \label{eqFmunu}
  \end{equation}
as is the curvature tensor of Riemannian geometry
 \begin{equation}
  -R^{\ \ \lambda}_{\mu\nu\ \sigma} V^\sigma = ([\nabla_\mu,\nabla_\nu] V )^\lambda.
% minus sign convention chosen to match of page 140 Weinberg GR
% Does not match Sean Carrol convention
  \label{eqR}
 \end{equation}
\par
%General relativity is based upon Riemannian geometry.

Unlike a traditional gauge theory,
the connection in Riemannian geometry
is not fundamental and can be derived
from a metric or an embedding.
Thanks to a proof by John Nash~\cite{56Nash01},
these two approaches are equivalent.
For every Riemannian metric, one can always find an isometric
embedding into a larger-dimensional flat space such that the
pull-back gives the same Riemannian metric.
These two equivalent ways to understand
the connection each have advantages.
The derivation of the connection from a fundamental metric
never uses dimensions outside the embedded manifold.
The derivation of the connection from an explicit embedding aids
in the understanding or visualization of the geometry.
\par
The parallels between gauge theory and Riemannian geometry lead to
our first question: Does a $U(n)$ gauge theory have an analogous
embedding from which one can derive the gauge field and can
visualize the geometry? The answer is yes.  In 1961, Narasimhan
and Ramanan~\cite{61Narasimhan,63Narasimhan} showed such an
embedding. They showed a surjective map from an embedding of a
${\bf C}^n$-vector bundle into a trivial ${\bf C}^N$-vector bundle
($N > n$) onto a standard $U(n)$ gauge theory, and they proved
that such an embedding exists for every $U(n)$ gauge field if $N
\ge (2\,d+1)n^3$ where $d$ is the number of space-time dimensions.
Since then, Atiyah~\cite{79Atiyah}; Corrigan Fairlie, Templeton,
and Goddard~\cite{Corrigan:1978ce}; Dubois-Violette and
Georgelin~\cite{Dubois-Violette:1979it}; Cahill and
Raghavan~\cite{Cahill:1993uq} all have used this map to represent
traditional gauge theories, as have others more
recently~\cite{Ikemori:1999yf,Valtancoli:2001gx}\@. Although not
identical, this map from an embedding to a gauge field is similar
to the corner variables of  Bars \cite{Bars:1978xy,Bars:1979qd}
and the degenerate subspace described by  Wilczek and Zee
\cite{Wilczek:1984dh}\@.
 In this paper, we study this gauge-field embedding applied to abelian
gauge theories.
\par
Previous work is discussed in
sections \ref{SecGaugeTheoryReview} through
%, \ref{SecChargeQuant}, \ref{SecRiemannianGeometry} and
\ref{SecGeometricalGaugeTheory}. First, this section reviews
traditional gauge theories. Next in section \ref{SecChargeQuant},
we show how the traditional gauge theory leads to the charge
quantization puzzle. Section \ref{SecRiemannianGeometry} reviews
how the metric and the connection are derived from an embedding.
In comparison to section \ref{SecRiemannianGeometry}, we explain
in section \ref{SecGeometricalGaugeTheory} the map from the
embedding onto a traditional gauge theory. Nearly all previous
work on embedded gauge theories focused on non-abelian gauge
theories; therefore, they took the unique coupling constant as a
given and they dealt with vector bundles of too large a dimension
to easily visualize.
\par
New material is covered in sections \ref{SecExamples} through
\ref{SecQuadraticGRTerm}. In contrast to previous work, we focused
our research on a very simple example: an $SO(2)$ gauge theory,
which is equivalent to a $U(1)$ gauge theory. In section
\ref{SecExamples}, we visualize this case as an
${\bf{R}}^2$-vector bundle embedded into a trivial
${\bf{R}}^3$-vector bundle. These simple $SO(2)$ examples forced
us to face questions not addressed in previous work: questions of
charge and coupling constants for different fields, and questions
of basis vector normalization. We found that to make the $SO(2)$
examples self-consistent, we needed to define a new type of gauge
theory that we call Riemannian gauge theory. In section
\ref{SecAction}, we introduce the Riemannian-gauge-theory action.
Our new action is built on the physical principle that a natural
description of nature should treat the four known forces with some
degree of symmetry. In this action, the gauge covariant derivative
is derived from an embedding and not defined by its transformation
properties. In the total space, the action treats all four forces
equivalently. Distinct forces arise after taking the projection
that defines the various fiber bundles. With an appropriate gauge
choice, our new action is nearly identical to the
traditional-gauge-theory action. Section \ref{SecChargeUniqueness}
emphasizes that the new action does not have the freedom to choose
distinct coupling constants for different matter-fields and
addresses the multiple $U(1)$ coupling constants in the standard
model. In section \ref{SecLargerSymmetry}, we observe that the
action has a larger $GL(n,{\bf{C}})$ local symmetry where the
gauge metric satisfies the metric compatibility condition. This
result is in sharp contrast to a traditional gauge theory of a
non-compact group. In the last section, we give our concluding
remarks.

\section{Charge Quantization History}
\label{SecChargeQuant}

Traditional gauge theories naturally quantize the charges of
non-abelian gauge groups but not those of abelian groups.
\par
In the abelian case, each field $\phi_i$ can be defined with a transformation
law specific to its own coupling constant $q_i$. The abelian field-strength
tensor (\ref{eqFmunu}) is a linear function of the coupling
constant, $F(q_1 + q_2,A)=F(q_1,A)+F(q_2,A)$, and
of the gauge fields, $F(q,A_1+A_2)= F(q,A_1)+F(q,A_2)$.
The field-strength term $\alpha\,F_{\mu \nu} F^{\mu \nu}$
in the action can be multiplied by any constant $\alpha$.
Because of the field-strength tensor's linearity and
the arbitrariness of $\alpha$, the covariant derivative of
any field $\phi_i$ can be used in eq.(\ref{eqFmunu})
to define $F_{\mu \nu}$.
The field-strength tensor is strictly gauge invariant
regardless of which coupling constant one chooses in the gauge-field
transformation rule (\ref{EqATransfRule}).
\par
In contrast, a non-abelian gauge group only allows one coupling constant $q$
for all non-trivially transforming fields.  The non-abelian field-strength
tensor is not linear in the coupling constant, nor is it linear in the gauge
field.  If different fields had different coupling constants,
then one would have to choose which coupling constant
to use in the gauge-field transformation rule
(\ref{EqATransfRule}) of the gauge fields in the field-strength tensor.
Also,  one would need to transform the gauge fields in the field-strength
tensor differently from the gauge fields in the covariant derivatives of the
matter fields.  This difference would violate gauge invariance.
So all fields transforming under a non-abelian gauge group
must have the same coupling constant.
\par
The coupling constant is different from the charge.
In the non-abelian case,
one will have different charges for matter-field multiplets proportional to
different eigenvectors of the group generators.  Simple groups will have
quantized charges.  In contrast, an abelian group has a continuum of
eigenvalues, and the matter fields may be proportional
to any corresponding eigenvector.
In the abelian case, no mathematical difference
distinguishes fields coupling with different coupling constants
from fields coupling with the same coupling constant
but with different charges.
\par
As far as we know, matter fields with an abelian
gauge group have charges that are integral multiples of $q_e/3$
where $q_e$ is the free-space charge of the electron. The
electric-charge quantization problem is the conflict between these
quantized electric charges and the unconstrained
continuum of allowable electric charges.
\par
Several physicists have offered explanations for the quantization
of electric charge.  Dirac~\cite{Dirac:1931kp} in 1931
showed that if magnetic monopoles exist, they would quantize
electric charge.
Georgi and Glashow~\cite{Georgi:1974sy} and Pati
and Salam~\cite{Pati:1974yy} explained the charges of the fermions
by associating the photon with a traceless generator of an
\(SU(n)\) unification group.
Others~\cite{Delbourgo:1972xb,Eguchi:1976db,
Alvarez-Gaume:1984ig,Babu:1989tq,Babu:1990ex,Foot:1993ui} have
exploited anomaly cancellation.

\section{Metric and Connection from an Embedding}
\label{SecRiemannianGeometry}

In Riemannian geometry~\cite{96Nakahara,97Lee,01Burago},
the covariant derivative may be defined
by properties of the connection and the metric,
or in-terms of a hypothetical flat embedding space.
In this section,
we present the Riemannian geometry framework of these two methods.
\par
First, we define the manifold with the tangent bundle.
A projection $\pi: E
\rightarrow M$ from a total space $E$
to a real, $d$-dimensional base manifold $M$
gives rise to a fiber bundle that is
the collection of vector spaces
$\pi^{-1}(x)={\bf R}^d$ for every space-time point
$x \in M$.  Because the fiber \(\pi^{-1}(x)\)
is a vector space, the fiber bundle is also
a vector bundle.
\par
The set of tangent vectors at a point $x$ is the set
of vectors that are tangent to all possible smooth,
differentiable curves that pass through $x$.
The fiber $\pi^{-1}(x)$ becomes a tangent space at
the point $p$ when one identifies directions
and lengths on the fiber
with directions and lengths on the base manifold.
This identification maps the set of tangent vectors at $x$
onto the vector space $\pi^{-1}(x)$.
Since the tangent fiber is a flat tangent plane,
only one set of basis vectors $t_\mu(x)$
at a fixed origin $O \in \pi^{-1}(x)$
is needed to fully define coordinates on the tangent space.
\par
A vector field ${V}$ is a tangent-space vector
that varies smoothly with the point \(p\) in $M$.
In a particular basis ${t}$,
%,also called a particular frame,
the vector field is given by
${V}=V^\mu {t}_\mu$ where the Greek indices run through the
dimensionality of the base manifold. The functions $V^\mu$ are
real-valued coefficients (i.e., not vectors)
of the basis vectors $t_\mu$.
A common choice for ${t}_\mu$ is the coordinate tangent
vectors, ${t}_\mu = \partial_\mu$\@.
\par
In addition to basis vectors,
we need dual basis vectors and an inner product.
The dual basis vector $\dualrep{t}^\mu$
is defined as the linear operator on the
basis vectors that returns the Kronecker delta:
$\dualrep{t}^\mu({t}_\nu)=\delta^\mu_\nu$.
The dual basis vectors combine with
the basis vectors to form a projection operator
\(P = t_\mu \, \dualrep{t}^\mu\)
which projects vectors onto the tangent space.
For a vector \(V\) in the tangent space,
the projection operator gives simply,
${P}({V}) = {t}_\mu
 \dualrep{t}^\mu( V^\alpha {t}_\alpha)={t}_\mu V^\alpha
 \delta_\alpha^\mu={t}_\mu V^\mu={V}$.
The inner product $ \langle V, W \rangle$
measures the length or the point-wise overlap
of two vector fields;
the inner product of two tangent vectors
is the metric,
\(g_{\mu\nu} =  \langle t_\mu, t_\nu \rangle\)\@.
In a particular basis, the inner product
of two vectors is
$\langle V,W \rangle=\langle V^\mu t_\mu, W^\nu t_\nu \rangle=
 V^\mu W^\nu  \langle t_\mu,t_\nu \rangle= V^\mu W^\nu
g_{\mu\nu}$.
\par
Given our fiber bundle $\pi: E \rightarrow M $,
the metric on the total space $E$ will be
bundle-like~\cite{Reinhart:1954ab,Boyer:2003ab}\@.
A bundle-like metric will satisfy
two properties relevant to this paper.
First, a bundle-like metric can be made
block diagonal in fiber components and
the base-manifold components~\cite{Reinhart:1954ab}\@.
Second, a bundle-like metric will have
$\partial_j g_{\mu\nu}=0$ where $\partial_j$
denotes a derivative with respect to
a coordinate that is tangent to the fiber at a base-manifold point and
where $g_{\mu\nu}$ are the base-manifold components
of the total space metric~\cite{Boyer:2003ab}\@.
\par
We will now show two methods to define the
covariant derivative.  The first defines properties
of the connection
and the metric, and the second is from a flat embedding space.
\par
The covariant derivative $\nabla_X V$ can be defined as
a map from two vector fields $X$ and $V$
to a new vector field.
The map must be linear in $X$, linear in $V$,
and must satisfy the product rule~\cite{97Lee}\@.
The covariant derivative of a vector field $V$
in the direction
$X^\sigma {t}_\sigma$ is written as
 \begin{equation}
 \nabla_X (  V) \equiv  X^\sigma t_\alpha ( \delta^\alpha_\mu \partial_\sigma +
\Gamma^\alpha_{\sigma \mu}) V^\mu
\label{EqGRCovariantDerivativeDefinition}
 \end{equation}
where the Christoffel symbols $\Gamma^\alpha_{\sigma \mu}$ define the connection.
We further define the metric $g$ to satisfy $\nabla_X g \equiv 0$.
These requirements lead to a unique connection $\Gamma^\alpha_{\sigma\nu}$
for a given metric $g$,
$\Gamma^\alpha_{\sigma\nu}=\frac{1}{2} g^{\alpha \lambda}
(\partial_\sigma g_{\lambda \nu}
+ \partial_\nu g_{\sigma \lambda} -
\partial_\lambda g_{\sigma \nu})$.
\par
Another way to introduce a covariant
derivative is to embed the manifold $M$
in a Euclidean (or Lorentzian) space $\embedsp{M}$
of higher dimension,
$X: M \rightarrow \embedsp{M}$.
Vectors in $\embedsp{M}$ are compared in the derivative
by means of ordinary parallel transport in the embedding space.
Using the embedding space, the covariant derivative of a vector field $V$
in the direction ${t}_\sigma$ is:
 \begin{eqnarray}
 \nabla_\sigma(  V) &\equiv&   P ( \partial_\sigma
(V^\alpha
  t_\alpha)) \equiv  t_\alpha ( \delta^\alpha_\mu \partial_\sigma
+
\Gamma^\alpha_{\sigma \mu}) V^\mu
\label{EqGRCovariantDerivativeDefinitionEmbedding} \\
\nabla_\sigma(  V) &=&
P(  t_\alpha \partial_\sigma V^\alpha
 +   V^\alpha \partial_\sigma  t_\alpha)
  =   t_\alpha\, \left( \delta_\mu^\alpha \partial_\sigma +
  \dualrep{t}^\alpha( \partial_\sigma
   t_\mu )\right) V^\mu .
    \label{EqGRCovariantDerivativeExpression}
 \end{eqnarray}
The Christoffel symbols
 \begin{equation}
 \Gamma^\alpha_{\sigma \mu}
\equiv \dualrep{t}^\alpha ( \partial_\sigma  t_\mu)
 \label{EqGRChristoffelDefinition}
 \end{equation}
are the projection onto the tangent space
of the derivative of the tangent vectors $t_\mu$ in the direction
${t}_\sigma$.
\par
If one begins with the properties of the connection,
Nash~\cite{56Nash01} proved that an isometric
embedding $X$ is always
possible with $\embedsp{M} = {\bf R}^D$
as long as $D \ge d(3d+11)/2$.
The explicit tangent vectors are found by coordinate derivatives on
the coordinates of the embedding function $\embedvec t_{\mu}^{\ j} =
\frac{\partial}{\partial x^\mu } X^j(x)$ yielding a $d
\times D$-dimensional rectangular matrix.
The metric on $M$ is given by the
inner product of two basis vectors $\langle t_\mu ,
t_\nu \rangle=g_{\mu \nu}= \delta_{ij} \embedvec t_\mu^i \embedvec t_\nu^j$.
This explicit embedding aids understanding and visualization.
\par
In Riemannian geometry, tangent-space basis vectors
can completely describe the intrinsic properties of the manifold
such as the connection, the metric, and the curvature tensors.
They can also describe quantities that are not
inherent to the manifold, such as the vector fields.

\section{Gauge Field from an Embedding}

\label{SecGeometricalGaugeTheory}
In section \ref{SecGaugeTheoryReview}, we reviewed traditional gauge theory
where the gauge fields are fundamental fields in the theory.
In this section, we review the way to derive
the gauge fields from an embedding built on the work of
Narasimhan and Ramanan~\cite{61Narasimhan,63Narasimhan}.
\par
We begin with a vector bundle given by a projection
$\pi_{F}: E_F \rightarrow M$ from a total
space $E_F$ to a space-time base
manifold $M$. The subscript $F$ denotes
the projection that creates the fiber.
The gauge vector bundle
is the collection of vector spaces
$\pi_G^{-1}(x)=F|_x$ for every point $x
\in M$. The fundamental matter field
determines the choice of the vector
space $F|_x$. For example, if $\phi$ is
a real, $n$-dimensional matter-field
multiplet, then the fiber is
$F|_x=\bf{R}^n$; and if $\phi$ is a
complex, $n$-dimensional field, then
the fiber is $F|_x=\bf{C}^n$. We shall
refer to the vector space $F|_x$ as the
\emph{fiber} or the \emph{gauge fiber}
in what follows.
\par
The geometry of the gauge vector bundle
differs from the tangent bundle because
directions or lengths on the fiber
$F|_x$  are not identified with
directions or lengths on the base
manifold.  Directions and lengths on
the gauge fiber are identified with
directions and lengths on the total space $E_F$;
however, these coordinates disappear under the
projection $\pi_F$.
Like the tangent fiber, the gauge fiber
is a flat plane, and so only one
basis-vector set $e_a(x)$ at a fixed
origin $O \in F|_x$ is needed to fully
define its coordinates.
\par
As in Riemannian geometry, a matter
field $\phi(x) \in F_p$ varies smoothly
with the point $x \in M$. In terms of a
local basis $e_a(x)$ on $F|_x$, the
matter field is given by
$\phi(x)=\phi^a(x) e_a(x)$ where the
Latin indices run through the
dimensionality of the gauge fiber. The
component \(\phi^a(x)\) of the matter
field $\phi(x)$ is a coefficient of a
basis vector $e_a(x)$ and is not
itself a vector.
The complex conjugate of the component $\phi^a$ is
denoted $\phi^{\bar a} \equiv \cmplxconj{\phi^a}$,
similarly \(e_{\bar a}\equiv \cmplxconj{e_a}\)\@.
The components \(\phi^{\bar a}\) are the coefficients
of the complex-conjugate basis vectors,
as in $\cmplxconj{\phi}=\phi^{\bar a} e_{\bar a}$\@.
\par
In addition to basis vectors,
we need dual basis vectors and an inner product.
Like the dual basis vectors of Riemannian geometry,
the dual basis vector $\dualrep{e}^a$ is defined as
the linear operator on the
basis vectors that returns the Kronecker delta:
$\dualrep{e}^a(e_b) = \delta^a_b$.
The dual basis vectors combine with the basis
vectors to form a projection operator $P=e_a \dualrep{e}^a$ which
projects vectors onto the gauge fiber.
For a matter field $\phi$ in the gauge fiber,
the projection operator gives simply, $P (\phi) = e_a
\dualrep{e}^a(\phi^c e_c)=e_a \phi^c \delta^a_c=e_c \phi^c= \phi$.
A complex vector space also has dual vectors
for the complex conjugate
of the basis vectors
$\dualrep{e}^{\bar a}( e_{\bar b } )
= \delta^{\bar a}_{\bar b}$\@.
By definition~\cite[p.~275]{96Nakahara},
one has
$\dualrep{e}^{\bar a}(e_b) =0$ and
$\dualrep{e}^{a}(e_{\bar b}) =0$.
\par
We use the notation $\langle \phi, \psi
\rangle$ to denote the inner product of
two matter vector fields, \(\phi\) and
\(\psi\)\@. In quantum mechanics, inner
products are used to compute lengths
and probability amplitudes. The inner
product of the basis vectors of the
fiber $F|_x$ is the gauge-fiber metric
$g_{\bar a b}=\langle e_a, e_b \rangle$,
which is distinguished by its Latin
indices from the metric \(g_{\mu \nu}\)
of the base manifold \(M\)\@.
The inner product of two complex matter vectors
is $\langle \phi, \psi \rangle =
\langle \phi^a e_a , \psi^b e_b \rangle
= {\phi^{\bar a}} \psi^b g_{\bar a b}$\@.
This inner product uses a hermitian metric,
\(\cmplxconj{g_{\bar a b}} = g_{\bar b a}\),
which by definition also satisfies
$g_{\bar a \bar b} = g_{a b} = 0$.
The gauge-fiber metric is defined so that
$g_{\bar a b} = g_{b \bar a}$
and $\cmplxconj{g_{\bar a b}} = g_{a \bar b}$\@.
The quantity \(g^{a \bar b}\)
is the inverse of the fiber metric,
$g^{a \bar b} g_{\bar b c} = \delta^a_c$\@.
The fiber metric and its inverse
can raise and lower indices:
\begin{equation}
  \phi_{\bar a} = g_{\bar a b} \phi^b
\qquad  \phi^{\bar a}=g^{\bar a b} \phi_b.
\end{equation}
Because $g^{a b}=0$, we cannot get $\phi^a$
by using $g^{a b}$ to raise an index.
Instead we write \(\phi^a = g^{a \bar b} \phi_{\bar b}\)\@.
\par
All previous research in gauge theories
with an explicit embedding space assumed orthonormal
basis vectors with a trivial gauge fiber metric.
We have generalized the
approach in the literature
to make manifest the symmetry with
Riemannian geometry.
\par
Given our fiber bundle $\pi_F: E_F \rightarrow M $,
the metric on the total space $E_F$ will be
bundle-like~\cite{Reinhart:1954ab,Boyer:2003ab}\@.
A bundle-like metric will satisfy
two properties relevant to this paper.
First, a bundle-like metric can be made
block diagonal in fiber components and
the base-manifold components~\cite{Reinhart:1954ab}\@.
Second, a bundle-like metric will have
$\partial_a g_{\mu\nu}=0$ where $\partial_a$
denotes a derivative with respect to
a coordinate that is tangent to the fiber of a base-manifold point  and
where $g_{\mu\nu}$ are the base-manifold components
of the total space metric~\cite{Boyer:2003ab}\@.
\par
Like the covariant derivative in Riemannian geometry,
the gauge-covariant derivative of a
matter multiplet $\phi$ in the direction $t_\sigma$ is the
projection onto the gauge fiber
of the derivative of the matter multiplet $\phi$
in the direction $t_\sigma$:
\begin{equation}
D_\sigma (\phi) \equiv P( \partial_\sigma (e_a \phi^a) ) \equiv
e_a \left(\delta ^a_b \partial_\sigma - i (A_\mu)^a_{\ b}\right)
\phi^b \label{EqGaugeCovariantDerivativeDefinition}
\end{equation}
\begin{equation}
D_\sigma (\phi) = P(  e_a \partial_\sigma \phi^a + \phi^a
\partial_\sigma e_a ) = e_a \left( \delta ^a_b
\partial_\sigma
+ \dualrep{e}^a(\partial_\sigma e_b ) \right) \phi^b ,
\label{EqGaugeCovariantDerivativeExpression}
\end{equation}
where the gauge field is
 \begin{equation}
 (A_\mu)^a_{\ b} =
i \dualrep{e}^a(\partial_\mu e_b).
 \label{EqGaugeFieldDefinition}
 \end{equation}
\par
We compare basis vectors
at different points of the manifold
by embedding the gauge fiber $F$ in a trivial,
real or complex Euclidean vector bundle
$M \times \embedsp F$
with a fiber $\embedsp F$ of
higher dimension: \( F \rightarrow \embedsp F\)\@.
In the spirit of Nash's embedding theorem~\cite{56Nash01},
Narasimhan and Ramanan~\cite{61Narasimhan,63Narasimhan}
showed that for any $U(n)$ or $SO(n)$ gauge field,
one can embed the gauge fiber $F$
in a trivial embedding fiber
$\embedsp F$ of dimension $N \ge (2\,d+1)n^3$\@.
The basis vectors of the fiber \(F \subset \embedsp F\)
now are the orthonormal vectors $\embedvec e^{\ j}_a$,
where the index \(j\) runs from 1 to \(N\)\@.
The $n$ basis vectors $e_a$ of \(F\)
are each $N$-vectors \(\embedvec e_a^{\ j}\)
in the trivial fiber and
span an $n$-dimensional subspace of \(\embedsp F\)\@.
The embedding space may now be used to express
the projection operator
\begin{equation}
P^j_{\ k} = \embedvec e_a^{\ j} \, \dualrep{\embedvec e}^a_{\ k}
\end{equation}
and the metric
\begin{equation}
g_{\bar a b} =
\cmplxconj{ \embedvec e^{\ k}_a} \,
\embedvec e^{\ j}_b \, \delta_{\bar k j}
= \embedvec e_{\bar a j} \,
\embedvec e^{\ j}_b
= \sum_{j=1}^N \,
\cmplxconj{\embedvec e_{aj}} \, \embedvec e_{bj},
\label{EqGaugeFiberMetric}
\end{equation}
in which the bar denotes
ordinary complex conjugation.
The quantity \(\delta_{\bar j k}\) may be
interpreted as a hermitian metric
on a complex Euclidean embedding fiber.
\par
Narasimhan and Ramanan based their
embedding theorem on orthonormal basis vectors
$ \langle e_a, e_b \rangle = \delta_{\bar jk} \,
\cmplxconj{\embedvec{e}_a^{\ j} } \, \embedvec e_b^{\ k}
= \delta_{\bar a b}$
spanning an $n$-dimensional subspace of an
$N$-dimensional embedding space.
Any two choices of real (complex)
orthonormal basis vectors may be related by
an orthogonal (unitary) transformation.
Gauge invariance \emph{is} this arbitrariness in the choice of basis vectors.
For real or complex orthonormal basis vectors,
Narasimhan and Ramanan showed that the resulting connection and
matter fields have an $SO(n)$ or $U(n)$ gauge symmetry.
The somewhat artificial constraint
$\dualrep{\embedvec e}^a_{\ j}
\partial_\mu \embedvec e_a^{\ j} = 0$
leads to an $SU(n)$ gauge group.
\par
The embedding
$\embedvec{e}: F \rightarrow \embedsp{F}$
is not unique,
and so we do not provide a general map
$(A_\mu)^a_{\ b} \rightarrow \embedvec e_a^{\ j}$
from the gauge field
to the embedded basis vectors.
The inverse map (\ref{EqGaugeFieldDefinition})
from the embedded basis vectors
to the gauge field is
\begin{equation}
(A_\mu)^a_{\ b} =
i \dualrep{\embedvec{e}}^a_{\ j}
\partial_\mu \embedvec e_b^{\ j}.
\label{EqGaugeFieldDefinitionEmbedding}
\end{equation}
Narasimhan and Ramanan have shown that
this formula for the gauge field
$(A_\mu)^a_{\ b}$ in terms of
embedded orthonormal basis vectors
is possible for every
$SO(n)$, $U(n)$, or $SU(n)$ gauge field.
The purpose of the embedding is to clarify the geometry and
mathematics of the basis vectors.
\par
All the basis vectors $e_a$ in the remainder of the paper should be
interpreted as existing in an unspecified larger dimensional embedding space.
To simplify the notation, we will drop the tilde.
\par
Several physicists~\cite{61Narasimhan,63Narasimhan,79Atiyah,Dubois-Violette:1979it,Corrigan:1978ce,Cahill:1993uq,Valtancoli:2001gx,Ikemori:1999yf}
have described traditional gauge theory
in terms of the orthonormal basis vectors
of a gauge fiber.
Next, we will consider some examples of an $SO(2)$
gauge theory and the resulting complications
relevant to the charge quantization puzzle.

\section{Visualizing Embedded Gauge Geometry}
\label{SecExamples}

\par
Embedding a vector bundle into a trivial vector bundle
enables a concrete visual understanding of the
geometry.  Example orthonormal basis vectors for the
mapping (\ref{EqGaugeFieldDefinitionEmbedding}) of Narasimhan and Ramanan
exist in previous publications.
Cahill and Raghavan~\cite{Cahill:1993uq} gave basis vectors
that represent a $U(1)$ plane wave.
Examples of instantons can be found in
refs.~\cite{79Atiyah,Ikemori:1999yf,Valtancoli:2001gx}
and references therein.  Self-dual examples
can be found in refs.~\cite{Corrigan:1978ce,Dubois-Violette:1979it}.
\par
At the end of this paper, we have two tables of
gauge-fiber embedding examples that map onto familiar $U(1)$
gauge fields.
In table \ref{TableU1FieldStrengths},
we give examples of ${\bf C}^1$ fibers embedded into  trivial ${\bf C}^2$
or ${\bf C}^3$ fibers.
In table \ref{TableSO2FieldStrengths},
we give examples of ${\bf R}^2$ fibers
embedded into a trivial ${\bf R}^3$ fiber.
The basis vectors in tables~\ref{TableU1FieldStrengths} and
\ref{TableSO2FieldStrengths} are orthonormal and map
onto a traditional, $U(1)$ gauge theory.
With the exception of the plane wave in eq.~(\ref{EqyPolarizedPlaneWaveU1}),
the examples shown here have never been published.

\par
Because the map from the embedding to the gauge
field is not invertable, no
unique map from a gauge field to the embedding exists.
We found these examples by trial and error.
The free parameters in the tables do not change
the gauge field or the field strength in any way.
These extra parameters highlight the many-to-one relationship
between the embeddings and the gauge fields.

 \FIGURE[t]{
 \centerline{\epsfig{figure=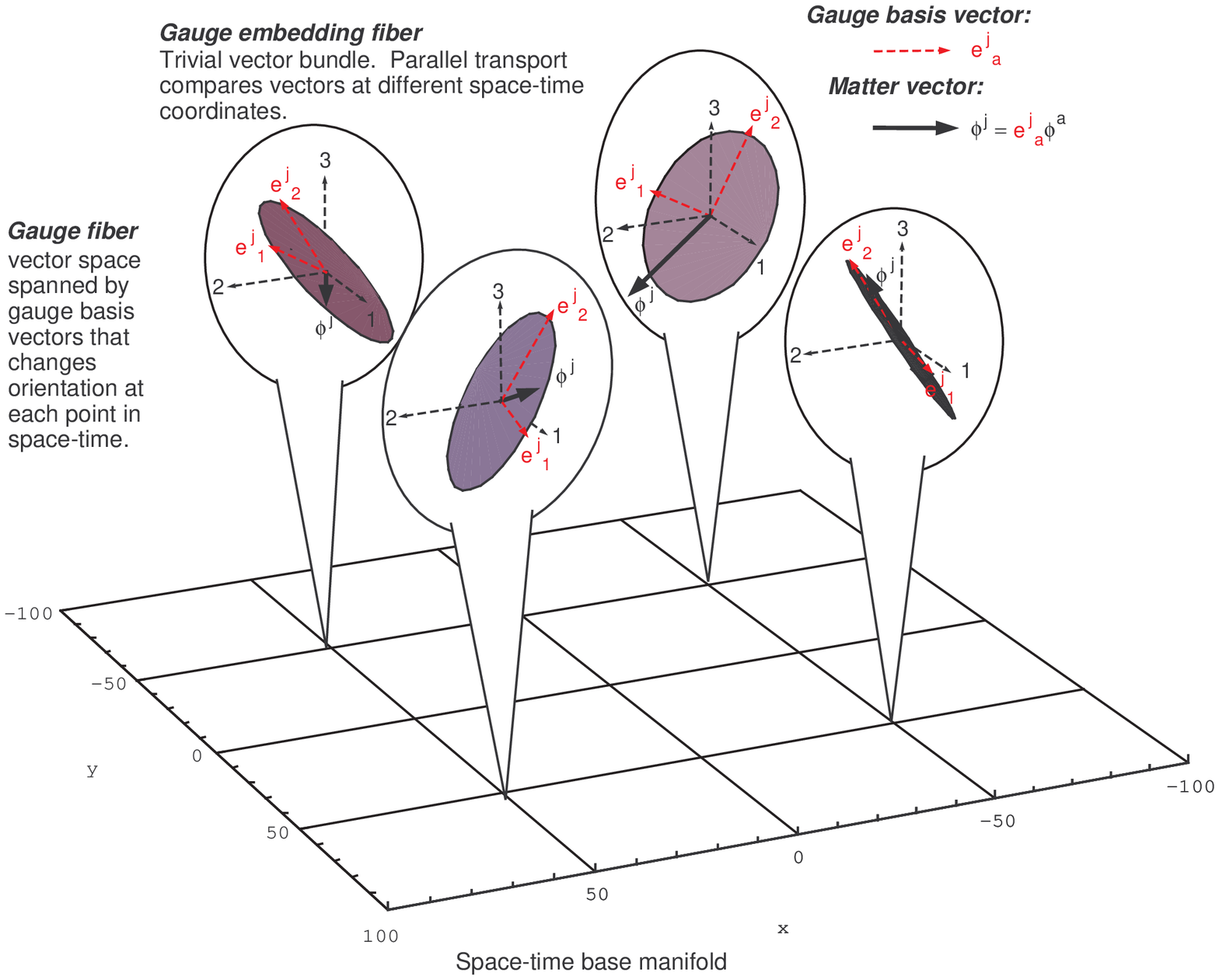,
angle=0,width=4.8in}}
 \caption{ An ${\bf R}^2$-vector bundle
embedded in a trivial ${\bf R}^3$-vector bundle.
The choice of gauge determines the dotted-red
basis vectors of the gauge fiber.
The thick black vectors of varying length represent
a real scalar matter field,
which can be interpreted as a wavefunction.}
 \label{FigBasisVectorIntro}
 }

\FIGURE{
%\begin{figure}
  \centerline{\epsfig{file=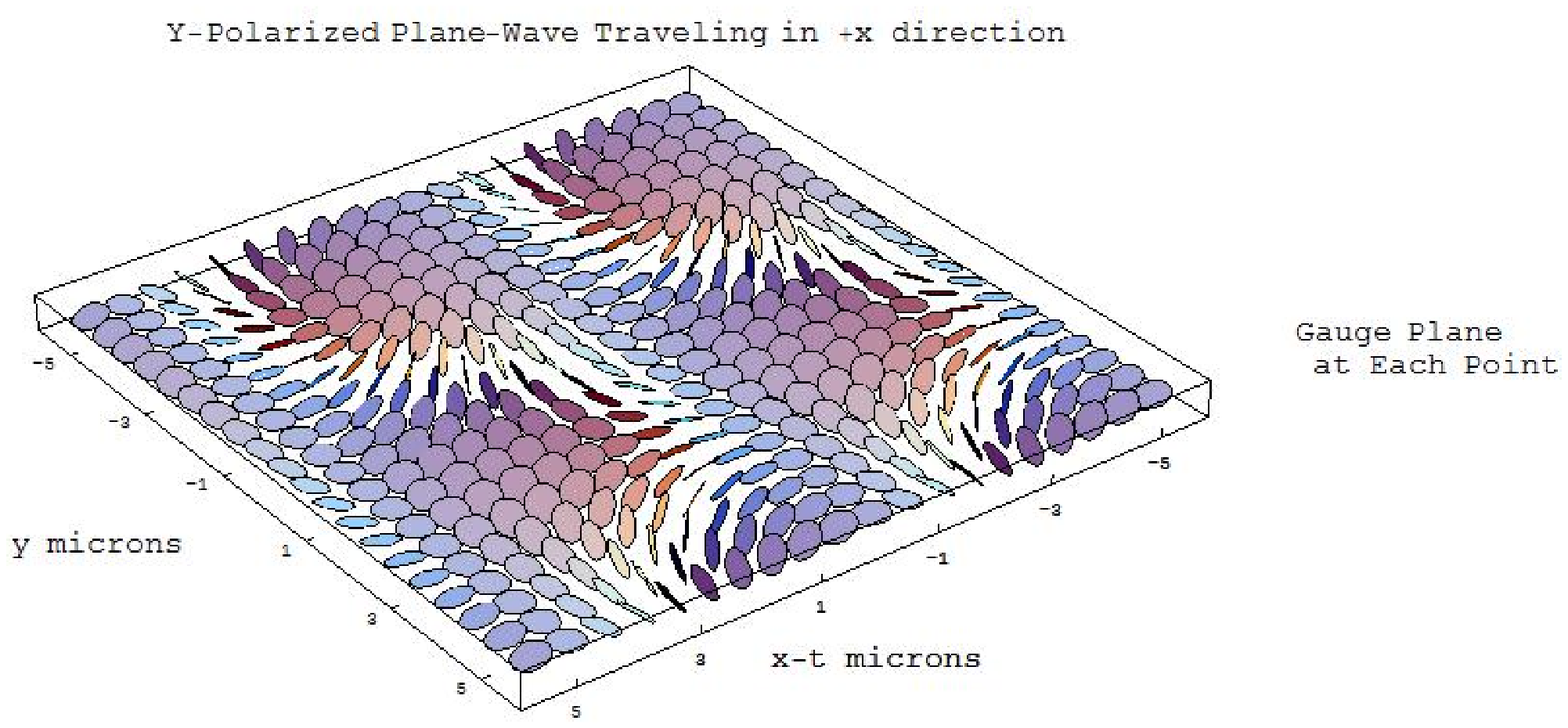,
angle=0,width=5.8in}}
  \caption{$\hat{y}$-polarized plane wave (\ref{EqPlaneWaveSO2})
  propagating in the positive $\hat{x}$
      direction.  The figure corresponds to radiation with
      $\lambda=10.6\ \mu\rm{m}$ and intensity of
      $\approx 6.4$ Watts/mm$^2$.
      }
  \label{Fig-YPolarizedPlaneWave}
%\end{figure}
}

\FIGURE{
%\begin{figure}
\centerline{\epsfig{figure=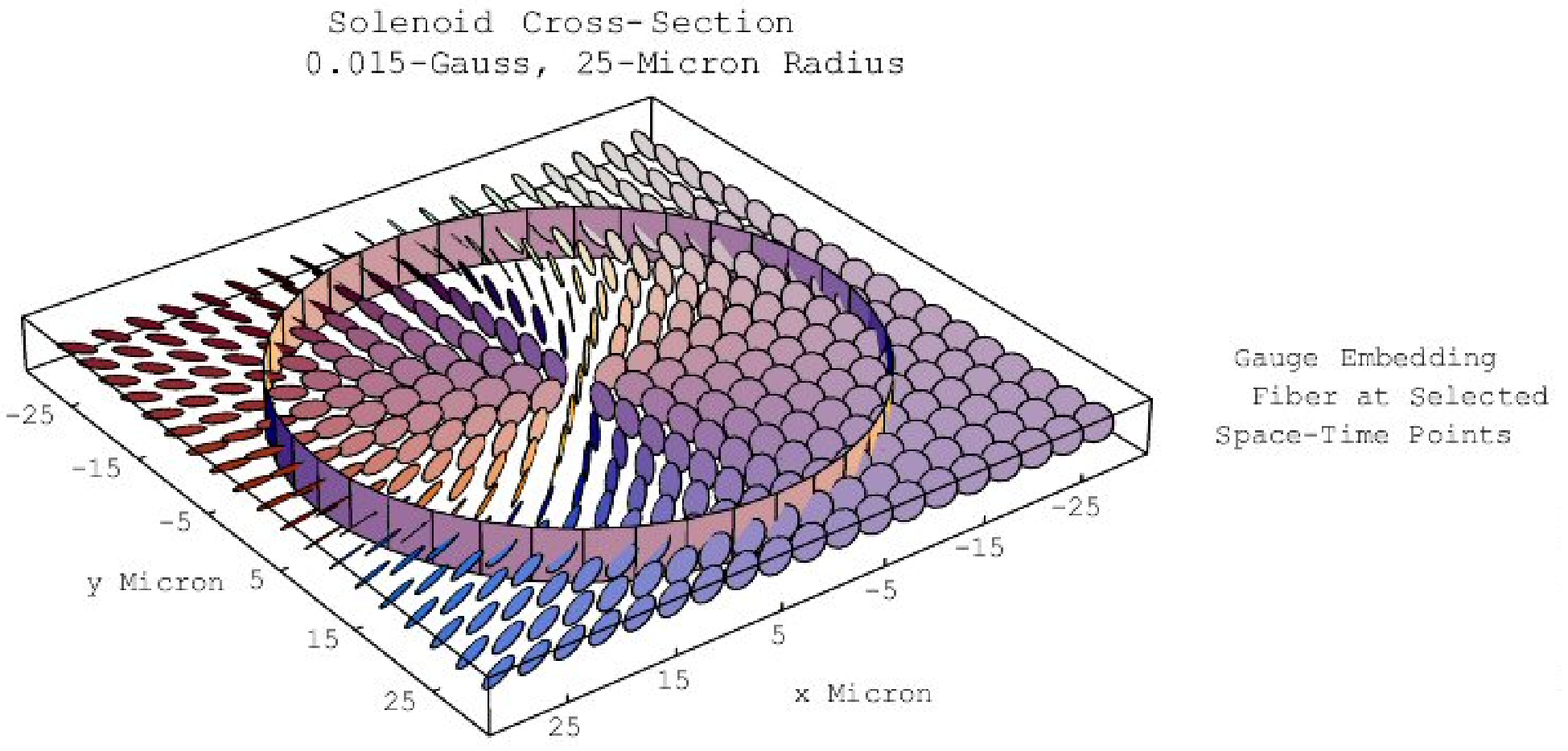,
angle=0,width=6.5in}}
\centerline{\epsfig{figure=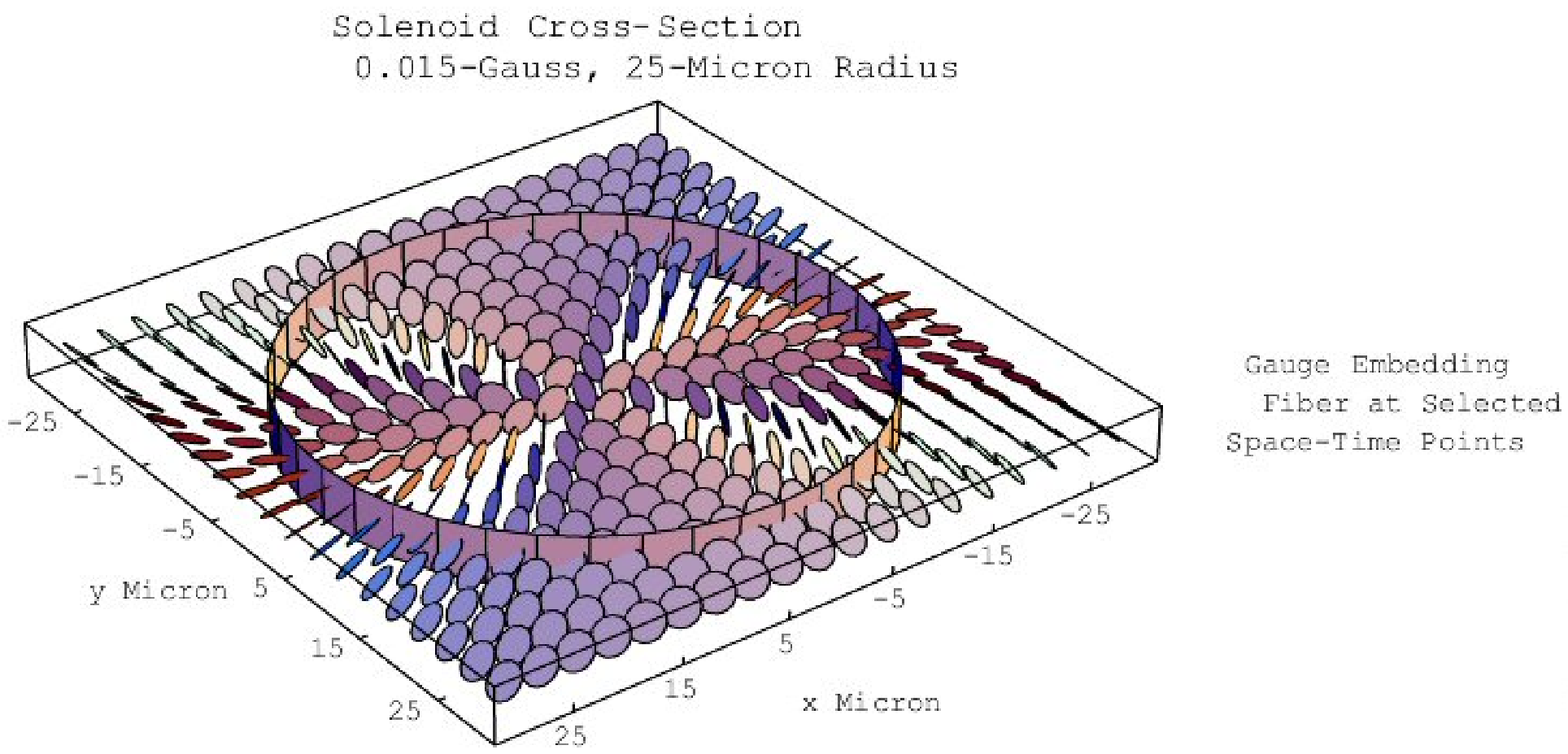,
angle=0,width=6.5in}} \caption{ \label{Fig-Solenoid-n1} The
cross-section of a solenoid of radius $R=25\,\mu\rm{m}$, with a
magnetic field strength of $B_o = 0.015\, \rm{Gauss}$. The diagram
shows eqs.~(\ref{EqInsideSolenoidSO2}) and
(\ref{EqOutsideSolenoidSO2}) for two vacuum topologies, $n=1$ in
the top figure and $n=2$ in the bottom figure. The cylinder
represents the solenoid boundary.}
%\end{figure}
}

\par
Figs.~\ref{Fig-YPolarizedPlaneWave}-\ref{FigTwoFieldsOnOneGeometry}
visually represent the $SO(2)$ examples in table
\ref{TableSO2FieldStrengths}\@. Fig.~\ref{FigBasisVectorIntro}
shows how to interpret
figs.~\ref{Fig-YPolarizedPlaneWave}-\ref{FigTwoFieldsOnOneGeometry}\@.
%UPDATE (add figure numbers, possibly adjust above).
Figs.~  will highlight how the gauge geometry of a constant
magnetic field curves the trajectory of a matter field of fixed
energy.

% Fig -- describes one example of how the curvature,
%  which is the ratio of the angle change in a vector by parallel transport of
% a vector around a loop, determines the radius of curvature of an
% electron in a magnetic field.

The bubbles in fig.~\ref{FigBasisVectorIntro} show the trivial embedding fiber ${\bf R}^3$
at periodic $x$ and $y$ coordinates.
The bubbles and the axis of the trivial embedding fiber,
shown in fig.~\ref{FigBasisVectorIntro},
are suppressed in the subsequent figures.
The gauge-fiber basis vectors,  $e^j_{\ 1}$
and $e^j_{\ 2}$, are shown as dotted, red
vectors that form right angles.
The disks represent the gauge-invariant, ${\bf R}^2$ subspace of the gauge fiber.
Geometrically, the electromagnetic field is the
changing orientation of the gauge fiber in the trivial, embedding fiber at
different space-time points.
The solid, black, vectors on the gauge fiber are the
matter fields, which can be interpreted as
a single-particle wavefunction.
\par
% Plane wave
Fig.~\ref{Fig-YPolarizedPlaneWave} shows a $\hat y$-polarized
plane wave (\ref{EqPlaneWaveSO2}) propagating in the $+\hat{x}$-direction.
The periodicity along the $x-t$ axis corresponds to a
wavelength of $\lambda=10.6\ \mu$m.
The changes in the gauge-fiber orientation
in the $y$-direction give
the $\hat{y}$-polarization and the intensity of $6.4$ Watts/mm$^2$\@.
\par
% Solenoid
%  - many-to-one map
%  - Topology
Fig.~\ref{Fig-Solenoid-n1} shows the cross-section of a solenoid
(\ref{EqInsideSolenoidSO2} and \ref{EqOutsideSolenoidSO2})
with a 25-$\mu$m radius.  Inside the cylinder,
the $z$-directed magnetic field is $0.015$ Gauss.  Outside
the cylinder, the magnetic field is zero.  The map
(\ref{EqGaugeFieldDefinitionEmbedding}) is surjective;
the two diagrams show two different embeddings that
give rise to the same magnetic field.  Many of
the examples in tables \ref{TableU1FieldStrengths} and \ref{TableSO2FieldStrengths}
have free parameters that give different embeddings but the
same gauge fields and the same field strengths.  The two diagrams
in fig.~\ref{Fig-Solenoid-n1} correspond to $n=1$ and $n=2$ topology
for the vacuum outside the solenoid.
\par
% Constant magnetic field
%  - Gauge Choice
%  - Wave function
%  - Negative rotation
\FIGURE{
   \centerline{\epsfig{file=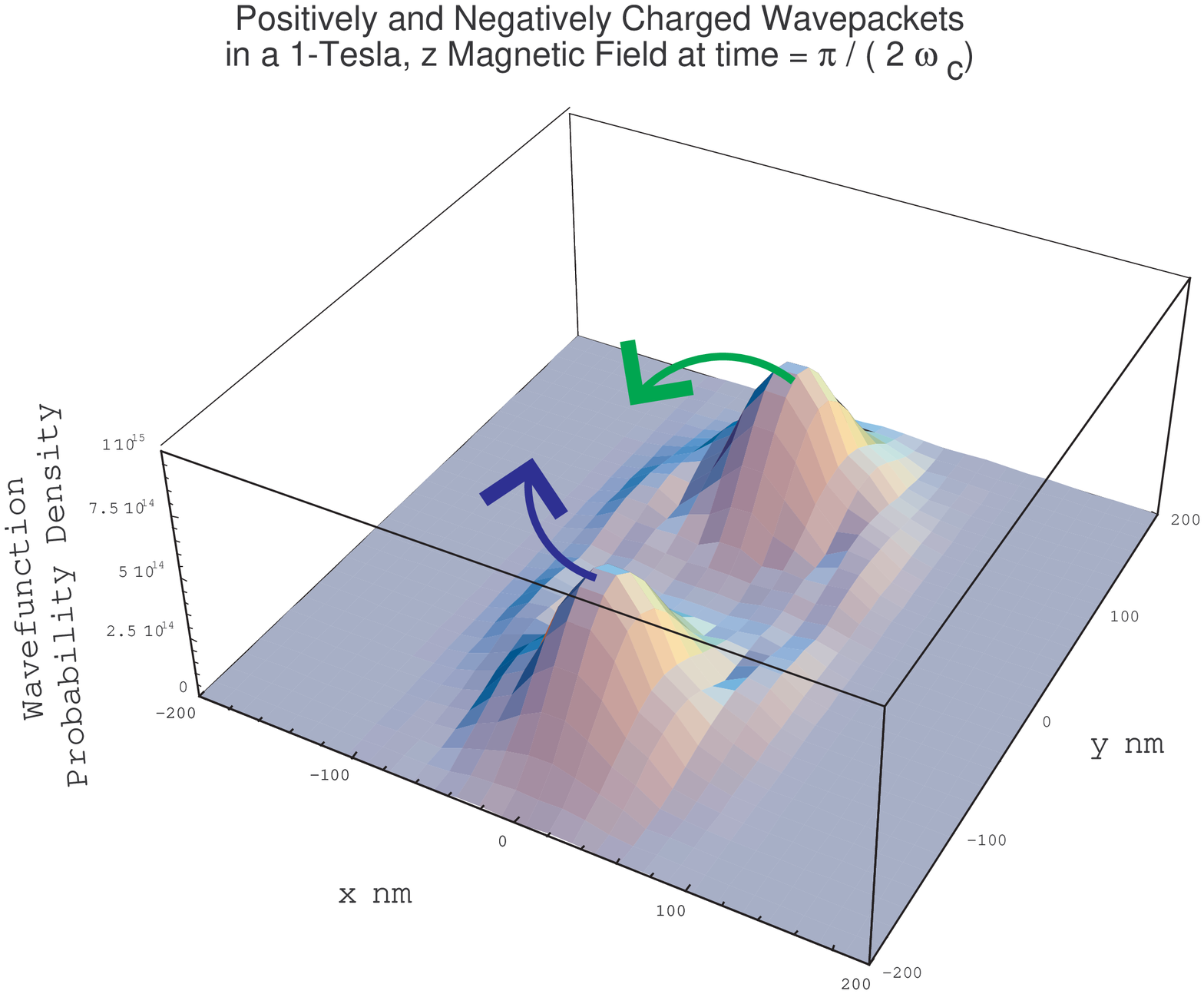,width=5in}}
 \caption{ The wave packet used in figures \ref{FigSO2MatterVectorPacket} through \ref{FigTwoFieldsOnOneGeometry}.
   \label{FigCyclotronSetup}} }
Figures \ref{FigSO2MatterVectorPacket} through \ref{FigTwoFieldsOnOneGeometry}
show positive and negative wave packets at various times during a cyclotron orbit
in the presence of a background, constant, z-directed magnetic field.
The wave packets used for these figures, shown in fig.~\ref{FigCyclotronSetup},
have an energy expectation value near the second Landau level;
they have the mass and apparent charge of an electron;
and their cyclotron orbit has a radius of about 50 nanometers.

\FIGURE{
%\begin{figure}
  \centerline{\epsfig{file=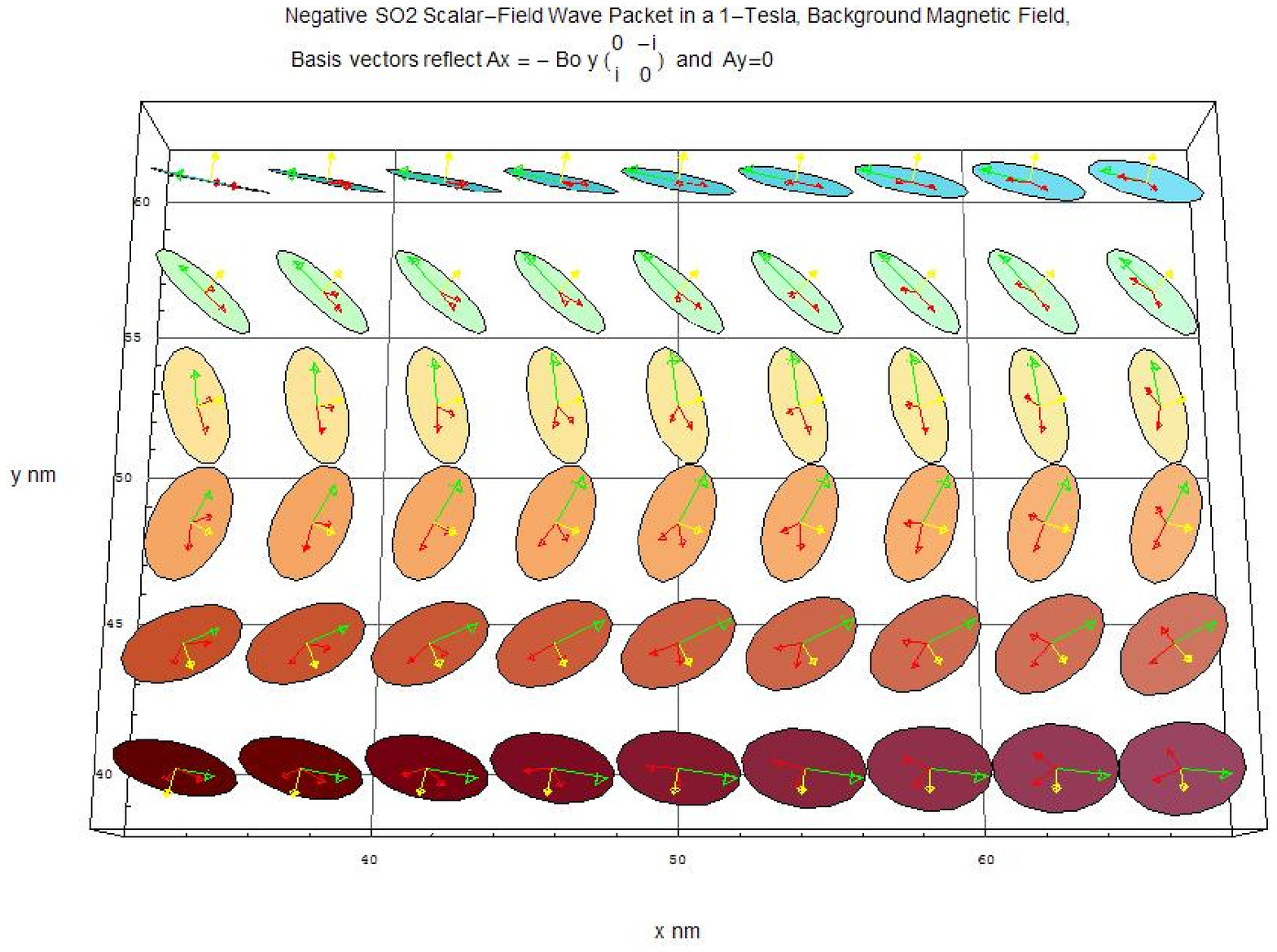,width=6.5in}}
  \centerline{\epsfig{file=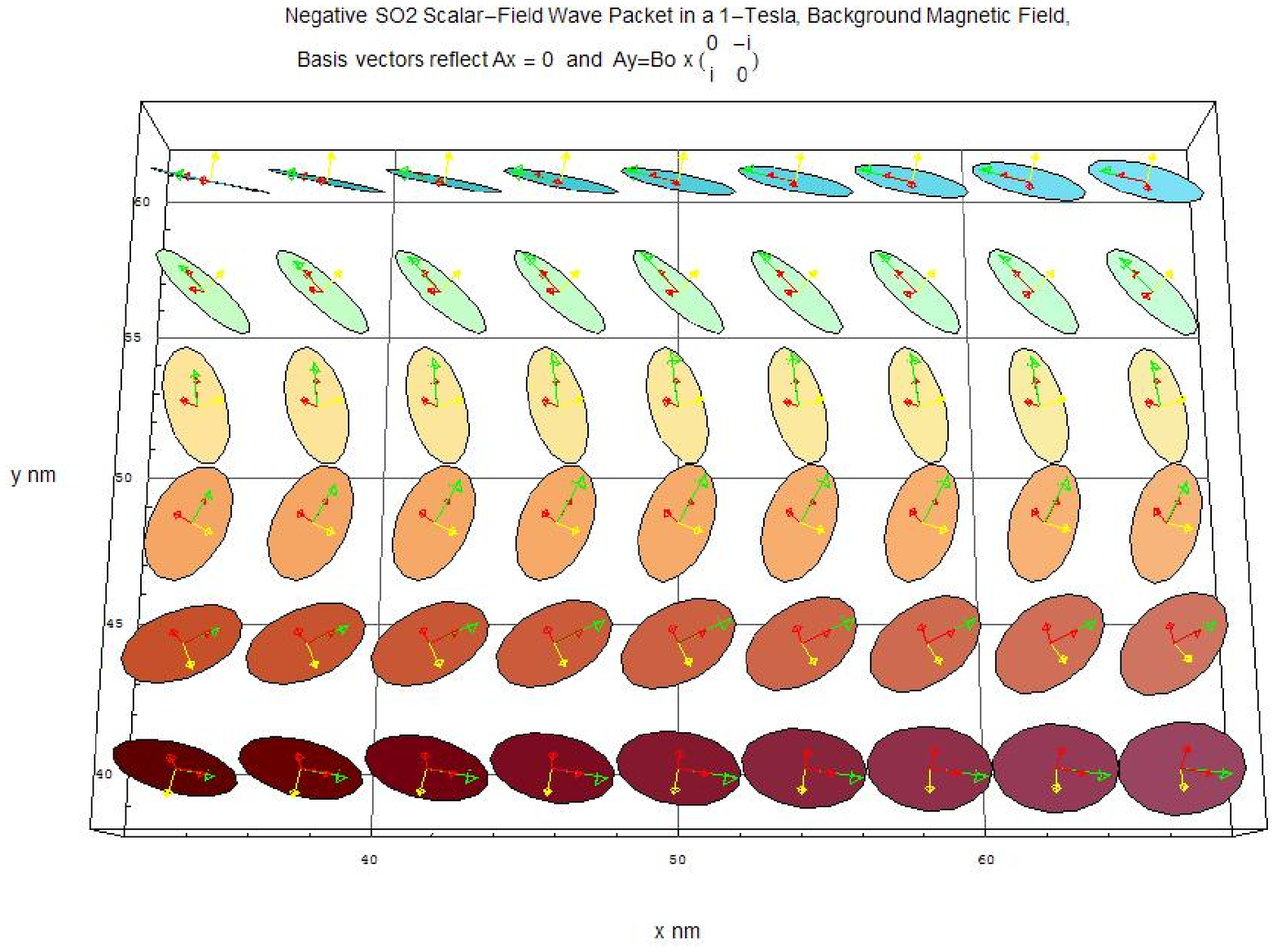,width=6.5in}}
  \caption{ The green vectors are a wave packet in a 1-Tesla magnetic field
  orbiting the point ($x=0$, $y=50$ nm) in a counterclockwise direction.
  The wave packet has energy centered near the second Landau level.
%  The red vectors at right angles are the basis vectors for the gauge fiber.
  The light-yellow vectors show the normal to the gauge fiber.
  The two diagrams show two different gauge choices. }
  \label{FigSO2MatterVectorPacket}
%\end{figure}
}

Fig.~\ref{FigSO2MatterVectorPacket} shows a constant $1$-Tesla
magnetic field with a negatively charged wave packet orbiting
counterclockwise about the point $(x=0, y=50\ {\rm{nm}})$.
The matter field is shown as a green vector.  To make clear the
orientation of the gauge fiber, we have added a light, yellow
vector indicating the normal of the gauge fiber.   The two figures
show two choices of gauge. In the top figure, the basis vectors
are related by parallel transport in the $y$-direction ($A_y=0$).
In the bottom figure, the basis vectors are related by parallel
transport in the $x$-direction ($A_x=0$). In both gauge choices,
the matter-field vectors remains fixed. Using the basis vectors in
the top figure, one can compare the matter vectors along the
$y$-direction and observe that they are rotating counterclockwise
in the momentum direction. For a positively charged wavefunction,
the vectors rotate clockwise in the momentum direction.
The rotating vectors may also be observed in the attached animation.
% ANIMATION LINE
%Show figure with both positive & negative wave packets on same diagram.
\par
These examples raised two questions: What is the role of the
basis-vector length? and What is the role of charge?
\par
In works on gauge theory with embedded basis vector, it is generally assumed that the basis vectors are orthonormal. From the
diagrams, one can see that the gauge-invariant physics is entirely
captured by the orientation of the gauge fiber, in this case the
${\bf{R}}^2$-subspace, and the matter-field vectors.  In
principle, these two objects may be equally well described by any
non-orthonormal basis vectors without changing the physics. The
resulting gauge fields are then proportional to generators of a
non-compact group.  However, a traditional gauge theory of a
non-compact group has completely different physics than a
traditional gauge theory of a compact group \cite{Cahill:1979qt}.
This apparent contradiction is resolved by defining the
Riemannian-gauge-theory action in section \ref{SecAction}.  The
resolution is then discussed more carefully in section
\ref{SecLargerSymmetry}.
\par
% *************************************
% BEGIN MAJOR EDITING SINCE REVISION 2
% *************************************
\FIGURE{
%\begin{figure}
(a)
\centerline{\epsfig{file=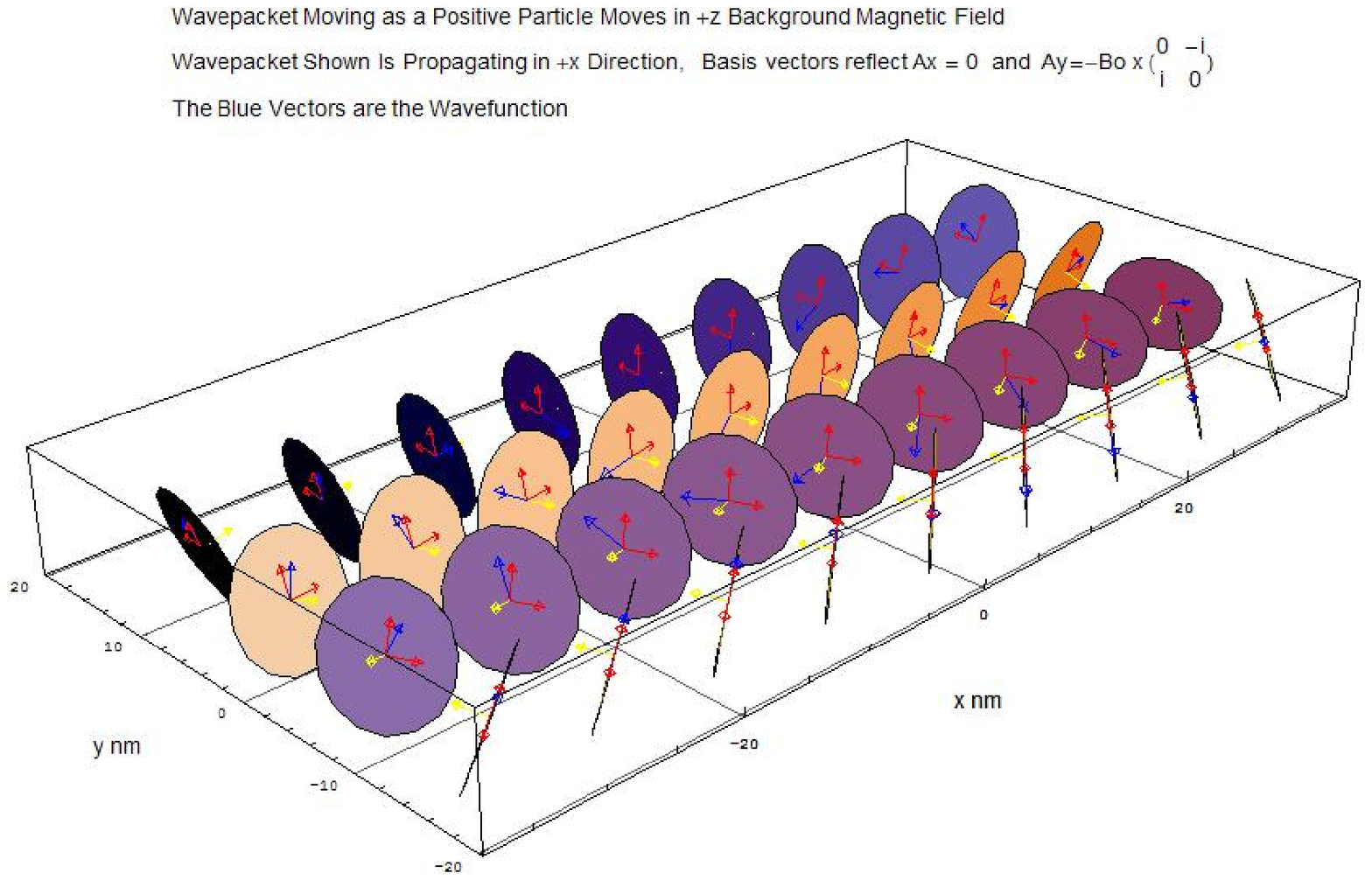,width=6.5in}}
(b)
\centerline{\epsfig{file=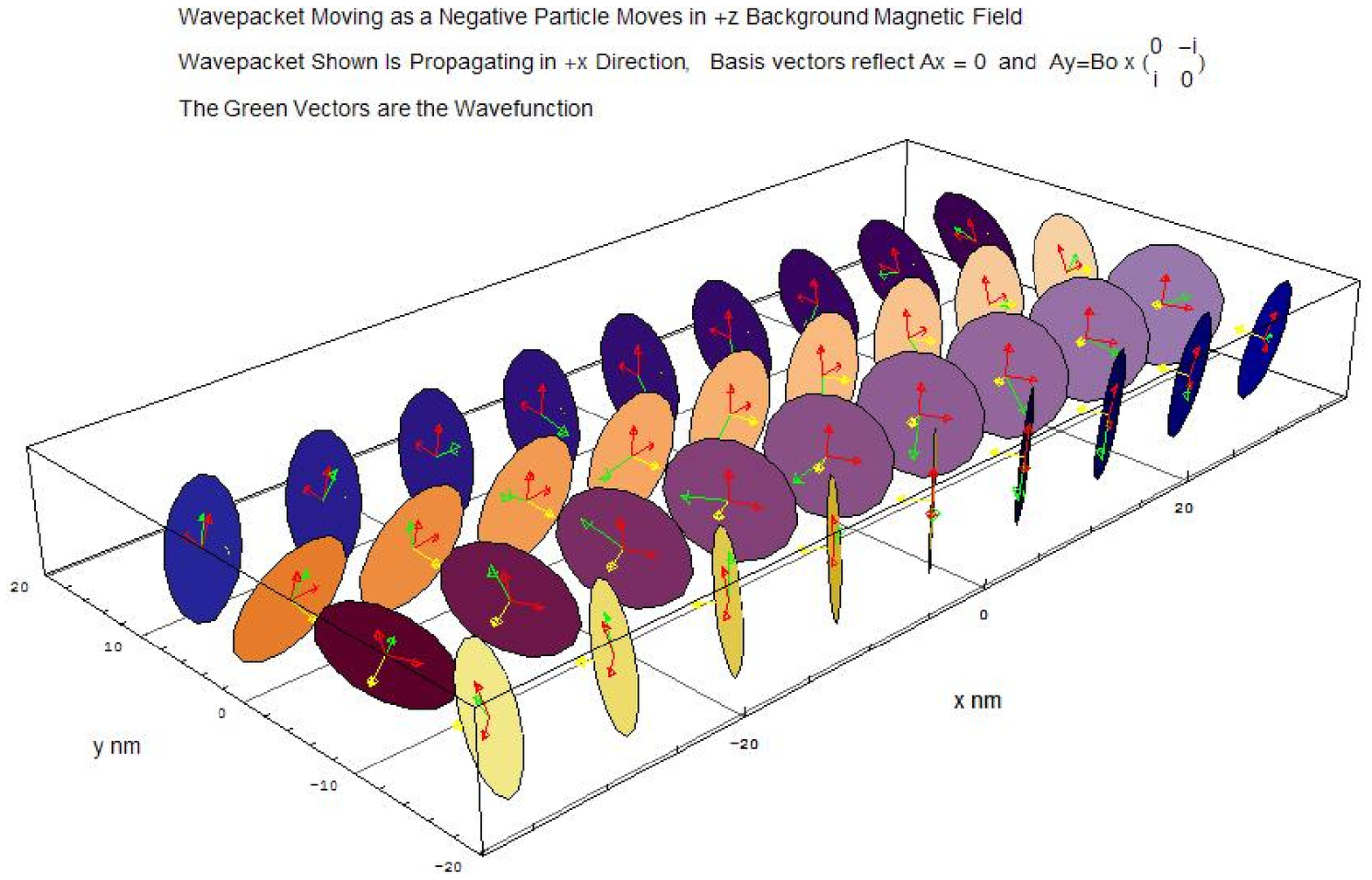,width=6.5in}}
 \caption{ The first method to incorporate two matter fields with different charges.
These two figures show two independent gauge fibers over the same space-time.
Part (a) shows a magnetic field with positive curvature, and part (b) shows a magnetic
field with negative curvature.
Without an external constraint, nothing restricts the
curvature of the two fibers to have any relation to each other.
\label{FigGeomChangingPair}} }
\par
Most authors writing on the geometry of gauge-fields
that use basis vectors have assumed a unique coupling constant or
only considered a single matter field.
In this paper, we consider multiple matter fields and explore how
the charge of each matter field can be varied.
We found two basic methods to add multiple matter fields.
\par
The first method establishes a separate basis vector for each matter field,
\begin{equation}
\phi=\phi^a e_a^{\ j} \ \ {\rm{and}} \ \ \psi=\psi^a E_a^{\ j}
\end{equation}
where $e_a^{\ j}$ and $E_b^{\ j}$ are two different sets of basis vectors that describe
two different gauge fibers at each space-time point.
Figure \ref{FigGeomChangingPair} depicts this first method to incorporate two matter fields.
The two matter fields are shown as  blue and green vectors.
The two parts of the figure show two independent gauge fibers over the same region of space-time.
Part (a) shows a magnetic field with positive curvature, and part (b) shows a magnetic
field with negative curvature.  By positive curvature, we mean that a vector parallel-transported
around a closed, clockwise loop in the x - y plane will be rotated clockwise in the same direction that a
vector parallel-transported on the surface of a sphere sphere is rotated.
\par
The matter field vectors of the wave packets in both parts (a) and (b) each rotate counterclockwise as one moves in the $\hat{k}$ direction.
Because the curvature in parts (a) and (b) are opposite, the wave packet in part (a) will curve as a positively charged particle, and
the wave packet in part (b) will curve as a negatively charged particle.
Because we have changed the trajectory of part (a) compared to part (b) by temporarily redefining the sign curvature associated with
the magnetic field, the
matter field in part (a) rotates in the opposite direction on the gauge fiber compared to the other positive matter fields in this paper.
\par
At this stage, there is no reason that $e_a^{\ j}$ and $E_a^{\ j}$ describe the same
electromagnetic field.  For example, $e_a^{\ j}$ could be a constant magnetic field given by eq.~(\ref{EqBConstRegionSO2})
and $E_a^{\ j}$ could be a plane wave given by eq.~(\ref{EqPlaneWaveSO2}).
There is nothing preventing one gauge fiber from representing the curvature of
an electric field and the second gauge fiber from representing a zero-curvature geometry indicating
no electric or magnetic field.
\par
In order that the fields $\phi$ and $\psi$ appear to be in the same electromagnetic field, we need to constrain the curvature of the gauge fiber defined by
by $e_a^{\ j}$ to be directly proportional to the curvature of the gauge fiber defined by $E_a^{\ j}$.
It is the curvature of the two geometries that must be proportional, not the basis vectors.
This method of incorporating multiple matter fields by placing each
field on a different gauge fiber would be equivalent in general relativity
to placing different particles on independent, different manifolds described by different metrics where every metric had a curvature
proportional to each other.
\par
\FIGURE{
%\begin{figure}
   \centerline{\epsfig{file=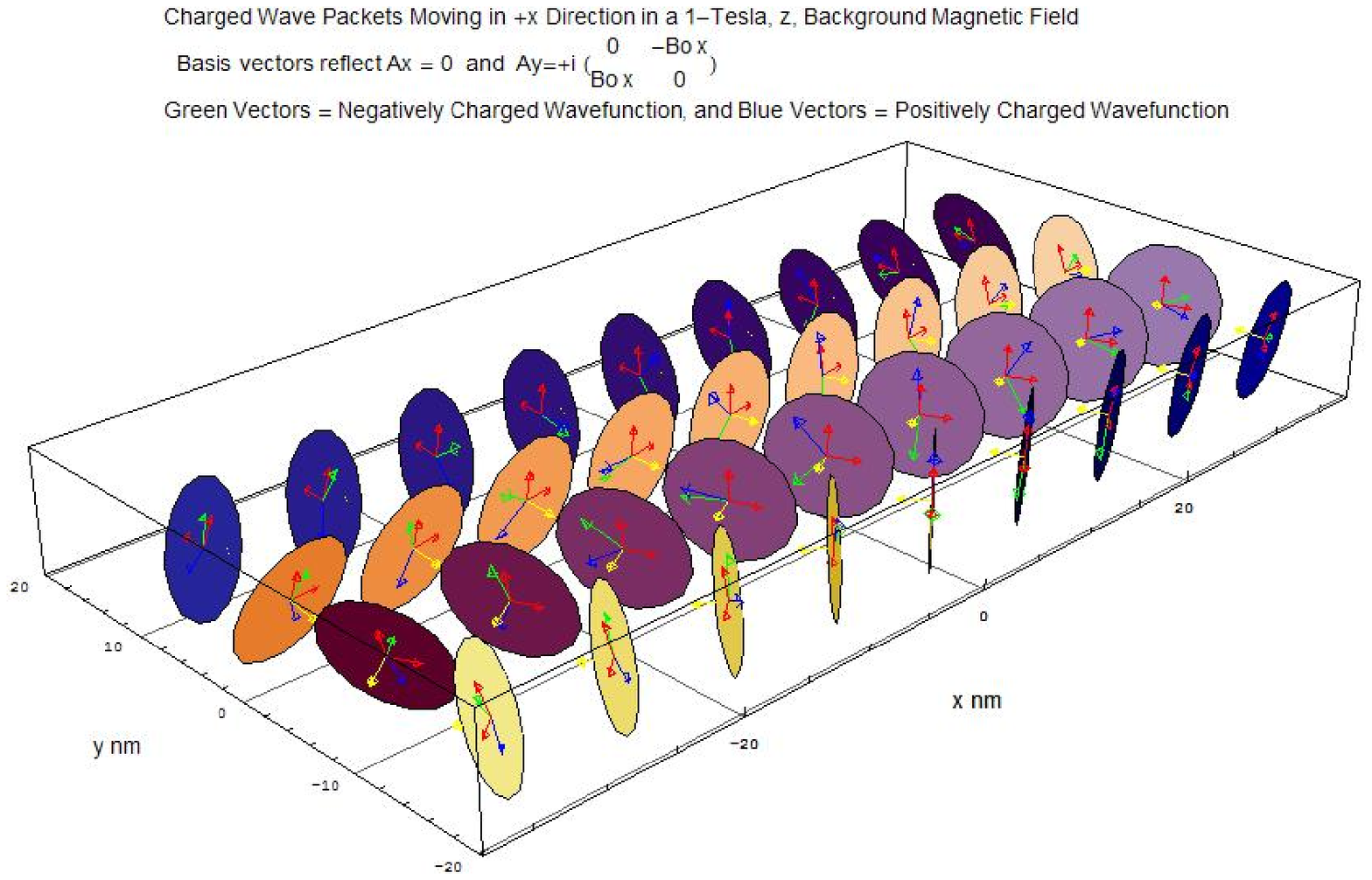,width=6.5in}}
 \caption{The second method to incorporate multiple matter fields places both matter-field vectors on
 the same gauge fiber.
  \label{FigTwoFieldsOnOneGeometry}}
%\end{figure}
}
The second method to incorporate multiple matter fields is to place the
two fields, $\phi$ and $\psi$, on the same background geometry.
Placing both fields on the same gauge geometry is accomplished
by writing both fields in terms of the same basis vectors:
\begin{equation}
\phi=\phi^a e_a^{\ j} \ \ {\rm{and}} \ \ \psi=\psi^a e_a^{\ j}.
\end{equation}
Figure \ref{FigTwoFieldsOnOneGeometry} depicts this second method.
The two matter fields each appear as separate blue and green vectors on the same gauge fiber.
\par
This method has the benefit that both matter fields automatically couple to the same
electric and magnetic fields.
Here, the covariant derivative (\ref{EqGaugeCovariantDerivativeExpression}) of both fields
will couple to the gauge field (\ref{EqGaugeFieldDefinition}) with the same coefficient.
The covariant derivative (\ref{EqGaugeCovariantDerivativeExpression})
has no free parameter that can be adjusted field by field.
\par
Positive and negative charges are possible by changing
the direction of rotation of the matter field on the gauge fiber.
A  counterclockwise or clockwise rotation as one moves in the $\hat{k}$ direction
determines if the wave packet is  negatively or positively  charged; therefore,
the two wave packets shown, although they are on the same background geometry, will be deflected in opposite
directions by the magnetic field.  Although we can manifest positive and negative charges,
we do not have the freedom to change the magnitude of the charge.
\par
This method of placing the two matter fields on a
single gauge fiber as in figure \ref{FigTwoFieldsOnOneGeometry}
is equivalent in general relativity to placing two particles on
a single manifold described by a single metric.
Because the particles share a common manifold,
they experience the same gravitational acceleration.
In gauge geometry, because the particles share a
common gauge fiber, they share a common charge magnitude.
Because of the connections to general relativity, in this paper, we
advocate this second method to incorporate multiple matter fields.
\par
The geometries and the figures shown in this section have led
us to reinterpret abelian gauge theory.
In a traditional, abelian gauge theory, each matter field can
couple with an arbitrary charge, and every matter field  automatically
sees the same background field.  Both of these features of traditional gauge
theory are problematic when the geometry is explicit.
In section \ref{SecAction}, we resolve this confusion by our definition
of the action of  Riemannian-gauge-theory.
The examples and figures in the present section
obey the equation of motion of this action.
The charge uniqueness of the new action
is revisited in section \ref{SecChargeUniqueness}.

%  WARNING: If we use the JHEP \TABLE command, then the figures and equations
%  inside the table are not referenced correctly
%\TABLE[t]{
\begin{table}[t]
  \centering
\begin{tabular}{|p{135pt}|p{290pt}|}
  % after \\: \hline or \cline{col1-col2} \cline{col3-col4} ...
  \hline
  \textbf{Description} &
  \textbf{ Basis Vector}  \\ \hline
  $\hat y$-Polarized Plane Wave
  \[A_y=2 A_o \cos^2 \frac{k}{2}(x-t)\]
%  \[F_{xy}=F_{ty}=A_o k \sin k(x-t)\]
  &
  \begin{equation}
 e^{j}_{\ 1} = \pmatrix{ e^{-i 2 A_o  y} \cos \frac{k(x-t)}{2 } \cr
 \sin \frac{k(x-t)}{2} \cr }
 \label{EqyPolarizedPlaneWaveU1}
 \end{equation}
  \\ \hline
  Circularly Polarized Wave
  \[ A_z=A_o \cos^2(k(x-t)) \]
    \[ A_y=A_o \sin^2(k(x-t)) \]&
  \begin{equation}
   e^{j}_{\ 1}=\left( \matrix{
 { f \cos(k (x - t) )  \exp(-i  \frac{A_o}{f^2} z) }\cr
 { f \sin(k (x - t) )  \exp(-i  \frac{A_o}{f^2} y) }\cr
 { \sqrt{1 - f^2 (\cos^2 k (x - t)  - \sin^2 k (x - t)  )} } \cr
} \right)
 \label{EqCircularlyPolarizedU1}
  \end{equation}
  where $f$ is any real value such that $0<f^2<1$. \\ \hline
  Finite Scalar Potential
  \[ A_t=\phi(\vec x) \]&
  \begin{equation}
  e^{j}_{\ 1}=\left( \matrix{ {
\sqrt{\frac{\phi(\vec{x})}{\kappa}}
\exp (-i\,  \kappa \, t) }\cr
 { \sqrt{1 - \frac{\phi(\vec{x})}{\kappa}} }\cr
 %{ 0 } \cr
} \right)
 \label{EqGenericScalarPotentialU1}
 \end{equation}
 where $0 < \rm{Min}(\phi(\vec{x})) < \rm{Max}(\phi(\vec{x})) < \kappa$.   \\ \hline
  Point Charge
  \[ A_t = \frac{q}{r} \]&
{  \begin{equation}
 e^{j}_{\ 1}(t,r)=\left( \matrix{
 {\sin(\kappa t) \exp( +i \frac{q}{2 r \kappa} \cot (\kappa t) ) }\cr
 {\cos(\kappa t) \exp( -i \frac{q}{2 r \kappa }\tan(\kappa t)   )}\cr
 %{ 0 } \cr
 } \right)
\label{EqCoulombBestU1}
\end{equation}}
  where $\kappa$ is any real value except zero.\\ \hline
  Solenoid:
  for $r<R$,
  \[ A_\phi = -\frac{B_o r^2}{2} \]
  \[ F_{r\phi}=B_o r \ \ F_{xy}=B_o  \]
  for $r>R$
  \[ A_\phi = -\frac{B_o R^2}{2} \ \  F_{r\phi}=0 \]
%  \[ \]
  &
  for $r < R$,
\begin{equation}
 e^{j}_{\ 1}(r,\phi)=\left( \matrix{
 {\sqrt{\frac{B_o r^2}{2 n}} \exp( - i n \phi) }\cr
 {\sqrt{1-\frac{B_o r^2}{2 n}} }\cr
 %{ 0 } \cr
 } \right),
\label{EqSolenoidU1Inside}
\end{equation}
and for, $r>R$,
\begin{equation}
 e^{j}_{\ 1}(r,\phi)=\left( \matrix{
 {\sqrt{\frac{B_o R^2}{2 n}} \exp( - i n \phi) }\cr
 {\sqrt{1-\frac{B_o R^2}{2 n}} }\cr
 %{ 0 } \cr
 } \right).
\label{EqSolenoidU1Outside}
\end{equation}
  In both cases, $n$ is any nonzero integer.
  \\ \hline
  Uniform Magnetic Field
  \[ A_x=-B_o y, \ \  F_{xy}=B_o \]
&  \begin{equation}
 e^{j}_{\ 1}(x,y)=\left( \matrix{
 {\sin(\kappa x) \exp( +i \frac{y\, B_o}{2 \kappa} \cot(\kappa x) ) }\cr
 {\cos(\kappa x) \exp( -i \frac{y\, B_o}{2 \kappa }\tan(\kappa x) )  }\cr
 } \right)
\label{EqUniformMagEverywhereU1}
\end{equation}
  where $\kappa$ is any real value except zero. \\ \hline
\end{tabular}
  \caption{\label{TableU1FieldStrengths}Familiar $U(1)$ field strengths
  represented by the basis vectors of a ${\bf C}^1$ fiber embedded in a trivial ${\bf C}^2$ fiber
  or a trivial ${\bf C}^3$ fiber.}
\end{table}
%}

%%%%%%%%%%%%%%%%%%%%%%%%%%%%%%%%%%%%%%%%%%%%%%%%%%%%%%%%%%%%
%  Next table
%%%%%%%%%%%%%%%%%%%%%%%%%%%%%%%%%%%%%%%%%%%%%%%%%%%%%%%%%%%%

%  WARNING: If we use the JHEP \TABLE command, then the figures and equations
%  inside the table are not referenced correctly

%\TABLE[t]{
\begin{table}[t]
  \centering
\begin{tabular}{|p{135pt}|p{290pt}|}
  % after \\: \hline or \cline{col1-col2} \cline{col3-col4} ...
  \hline
  \textbf{Description} &
  \textbf{Basis Vector}  \\ \hline
 Plane wave
 \[ A_y = A_o \cos (k(x-t)) \]
 (see fig.~\ref{Fig-YPolarizedPlaneWave})
 &
 \begin{eqnarray}
 {e^{j}_{\ 1}  = \left( \matrix{ {\cos A_o y  }\cos k
(x-t) \cr {\sin A_o y \cos k (x-t) }
\cr {-\sin k(x-t) }\cr }
\right)} \nonumber \\
{ e^{j}_{\ 2}  = \left( \matrix{ {
-\sin A_o y }\cr {\cos A_o y }\cr {0
}\cr } \right) } \label{EqPlaneWaveSO2}
\end{eqnarray}
 \\ \hline

 Solenoid

 For $r<R$,
  \[ A_\phi = -\frac{B_o r^2}{2} \]
  \[ F_{r\phi}=B_o r \]
  \[ F_{xy}=B_o \]
  For $r>R$
  \[ A_\phi = -\frac{B_o R^2}{2} \]
  \[ F_{r\phi}=F_{xy}=0 \]

   (see fig.~\ref{Fig-Solenoid-n1})
 &
 For $r<R$,
 \begin{equation}
\matrix{ { e^{j}_{\ 1}=\left(\matrix{
{\cos(n \phi)\ }\cr {\sin(n \phi)\
}\cr {0} \cr
      } \right)
} & { e^{j}_{\ 2}=\left(\matrix{
{\sin(n \phi)\ \frac{B_o\,r^2}{2\,n}
}\cr {-\cos(n \phi)\
\frac{B_o\,r^2}{2\,n} }\cr {\sqrt{1 -
(\frac{B_o\,r^2}{2\,n})^2}} \cr }
\right) } },
\label{EqInsideSolenoidSO2}
\end{equation}
 and for $r>R$,
\begin{equation}
\matrix{ { e^{j}_{\ 1}=\left(\matrix{
{\cos(n \phi)\ }\cr {\sin(n \phi)\
}\cr {0} \cr
      } \right)
} & { e^{j}_{\ 2}=\left(\matrix{
{\sin(n \phi)\ \frac{B_o\,R^2}{2\,n}
}\cr {-\cos(n \phi)\
\frac{B_o\,R^2}{2\,n} }\cr {\sqrt{1 -
(\frac{B_o\,R^2}{2\,n})^2}} \cr }
\right) } }.
\label{EqOutsideSolenoidSO2}
\end{equation}
In both cases, $n$ is any nonzero integer.
  \\ \hline

Constant Magnetic Field.

\[ A_y = B_o x \]

Where $|x|< a/ B_o$.

(see
fig.~\ref{FigSO2MatterVectorPacket})
&

\begin{eqnarray}
e^{j}_{\ 1}& =  &\left(\matrix{ { \cos
y \kappa }\cr { \sin y \kappa}\cr {0
}\cr
      } \right) \nonumber \\
e^{j}_{\ 2} & = &\left(\matrix{ {
-\frac{B_o x}{\kappa} \sin y \kappa
}\cr { \frac{B_o x}{\kappa} \cos y
\kappa}\cr {\sqrt{1-(\frac{B_o
x}{a})^2} }\cr
      } \right),
\label{EqBConstRegionSO2}
\end{eqnarray}
where $\kappa$ is any real value except zero. \\ \hline

\end{tabular}
  \caption{\label{TableSO2FieldStrengths}Familiar $U(1)$ field strengths
  represented by the basis vectors of an ${\bf R}^2$ fiber embedded in a trivial ${\bf R}^3$ fiber.}

\end{table}
%}

\section{The Riemannian Gauge Theory Action}

\label{SecAction}

We construct the action for Riemannian
gauge theory to be as similar as possible
to the action of general relativity.
We focus on writing the action in terms
of inner products on the gauge fiber,
which is equivalent to contracting
upper indices with lower indices.
\par
Both in general relativity and in
Riemannian gauge theory,
the action of the matter field
is formed by inner products on the
gauge fiber integrated over space-time,
here taken to be four dimensional.
For spinless bosons the action is
\begin{equation}
S_\phi = - \int d^4 x \sqrt{- g}\ \left[
\langle D_\mu \phi , D^\mu
\phi \rangle + \langle m
\phi, m \phi \rangle \right],
 \label{EqPhiAction}
\end{equation}
and for fermions it is
\begin{equation}
S_\psi = - \int d^4 x \sqrt{- g}\ \left[ \langle
 \psi , i\, \gamma^0 \,\gamma^\mu\,D_\mu \psi \rangle +
 \langle  \psi, m \, i \, \gamma^0
\psi \rangle \right].
\label{EqPsiAction}
% Matches notation 8.6.1 in Weinberg QFT Vol 1
\end{equation}
The factor $\sqrt{-g}$ is the square
root of $-1$ times the determinant of the metric
of the base manifold.
\par
Both in general relativity and in
Riemannian gauge theory,
the action of the gauge fields
measures the intrinsic Riemannian curvature.
In general relativity, the curvature tensor
(\ref{eqR}) represents the
change in a vector \(V\) due to
parallel transport around a loop.
If the infinitesimal parallelogram starts in the $dx^\mu$
direction and then continues in the $dx^\nu$ direction,
this change is
\begin{equation}
(V'-V)^\sigma =
dx^\mu\,dx^\nu\,R^{\ \ \sigma}_{\mu\nu\ \lambda}\,
V^\lambda.
\end{equation}
The tensor \(R\) is
the fundamental measure of intrinsic
curvature.
The action of general relativity
\begin{equation}
S_{\rm GR}=-\frac{1}{16\pi G}\int d^4x \sqrt{-g}\,
R_{\mu\nu\lambda\sigma} \, g^{\mu \lambda}
\, g^{\nu \sigma}
\end{equation}
is formed from the
contracted curvature tensor.
\par
In Riemannian gauge theory,
the field-strength tensor
(\ref{eqFmunu}) also represents the change
\begin{equation}
(\phi'-\phi)^a=i\,dx^\mu\, dx^\nu\,
{F_{\mu \nu}}^a_{\ b}\phi^b
\end{equation}
in a matter field \(\phi\) due to
parallel transport around a loop.
Therefore ${F_{\mu \nu}}^a_{\ b}$
measures the intrinsic curvature of the
gauge fiber.
Both $R$ and $F$ measure intrinsic curvature,
and their definitions (\ref{eqFmunu}) and (\ref{eqR})
are nearly identical.
\par
We wish to measure symmetrically the curvature of the gauge fiber
and the curvature of the base manifold.
To make the relationship concrete, we
consider the $(4+4+n)$-dimensional
total space $E_T$ before taking
the projection $\pi_T=\pi_F \circ \pi$ which
combines the projection $\pi_F$ for the
fiber bundle of the gauge theory and the
projection $\pi$ for the tangent bundle
of the base manifold.
We will choose the simplest total-space
action that after the projection
$\pi_T$ measures the curvature of
the base manifold and of the fiber
in a coordinate independent manner.
We will ignore all contributions to the action which vanish
under the projection $\pi_T$.

Given our fiber-bundle geometry, the metric of the total-space $E_T$ is
bundle-like\footnote{We are indebted to C.~Boyer for bringing
the bundle-like metrics conditions to our attention.}.
A bundle-like
metric $g$ on the total space satisfies $\partial_a g_{\mu\nu} = 0$
and $g_{a\mu}=g_{\mu a}=0$~\cite{Reinhart:1954ab,Boyer:2003ab},
where
the indices $a$, $b$, $c$ and $d$ still run over the fiber dimensions, and
the indices $\mu$, $\nu$, $\sigma$ and $\lambda$ still run over the
base manifold dimensions.
In the total space $E_T$, the field-strength tensor is
a sub-tensor of the total-space curvature tensor,
\begin{equation}
R_{\mu \nu \ b}^{\ \ a}= i\,{F_{\mu\nu}}^a_{\ b}
\ \ \ {\rm{where}}\ \ \matrix{\mu,\nu&:&1 \ldots 4 \cr
                        a, b&:& 5 \ldots 4+n \cr} .
\label{EqRFRelationshipship}
\end{equation}

This expression warrants a few notes of clarification:
This equation is not a definition of our choice;
rather, the equation points out that the
standard curvature tensor in the total space
has the gauge theory field-strength tensor as a sub-tensor.
The curvature tensor of the total space is a measure
of the base manifold curvature and the gauge fiber curvature.
The curvature of the base manifold is measured by
the curvature of the tangent bundle due to the
identification of directions and lengths on the tangent bundle
to directions and lengths on the base manifold.
The total space $E_T$ includes coordinates tangent to the
gauge fiber which disappear under the projection
$\pi_F$.
The coordinates and lengths of the gauge fiber are
identified with the coordinates and lengths of
the total space that are tangent to the fiber.
Therefore, the curvature of the gauge fiber bundle
is captured in the components (\ref{EqRFRelationshipship}) of
total-space curvature tensor.
\par
Does the total-space Ricci scalar measure the curvature in the gauge fiber?
Under the fiber-like metric conditions, the Ricci tensor and the curvature
scalar cannot be formed from the field strength
tensor \(R_{\mu \nu \ b}^{\ \ a}=iF_{\mu\nu\ b}^{\ \ a}\).
In effect, because the metric
components $g^{\mu a}$ vanish,
both the Ricci tensor \(R_{\mu \nu a b}\, g^{\mu a}\)
and the curvature scalar
$R_{\mu \nu a b}\, g^{\mu a} g^{\nu b}$ also vanish.

To include in the action a measure of
the gauge-fiber curvature, we must use
higher-order terms that do not appear in
general relativity.  After taking the $\pi_T$
projection, we find the term
\begin{equation}
S_F = \int d^4x \sqrt{-g}  \
\frac{-1}{2\,q^2} {F_{\mu \nu}}^a_{\ b}
{F^{\mu \nu}}^b_{\ a}
\label{EqGFAction}
\end{equation}
is the first gauge-invariant,
Lorentz-invariant, higher-order measure
of curvature of the gauge fiber that
does not vanish due to symmetries or
the block-diagonal total-space metric.
\par
The field-strength action term (\ref{EqGFAction})
comes from the total-space action term quadratic in the
curvature tensor.
When we project this term with $\pi_T$ onto the base
manifold, we find a new contribution to the action from
the base manifold curvature:
\begin{equation}
S_{\rm GR-new}= \int d^4x\, \sqrt{-g}\,\frac{+1}{2 q^2} \,
R^{\ \ \sigma}_{\mu\nu\ \lambda}
R^{\mu\nu\lambda}_{\ \ \ \sigma}.
\label{EqGRNew}
\end{equation}
The sign of this expression relative to eq.~(\ref{EqGFAction})
is set by eq.~(\ref{EqRFRelationshipship}).
If the coupling in eq.~(\ref{EqGRNew}) is of the same
order of magnitude of the other three forces,
this new term would be a very small
correction to the action of general relativity.
\par
The basis-independent action of Riemannian gauge theory is
\begin{equation}
{S}={S}_\phi + S_\psi + {S}_F + S_{\rm GR} + S_{\rm GR-new}.
\label{EqRGTAction}
\end{equation}
To preview the physics contained in eq.~(\ref{EqRGTAction}),
we perform a small thought experiment.
First, we temporarily neglect gravitational
contributions and choose an orthonormal basis
for the fiber $F$.  Next, Narisimhan and Ramanan showed
in an orthonormal basis, the gauge fields ${A_\mu}^a_{\ b}$
are $U(n)$ gauge fields.
Because the action is independent of the choice of basis,
this thought experiment shows the physics in eq.~(\ref{EqRGTAction})
is always that of a $U(n)$ gauge theory with a small
general-relativity correction.

\section{Uniqueness of the Coupling Constant}
\label{SecChargeUniqueness}

A traditional gauge theory of a compact simple non-abelian group has a unique
coupling constant and discrete charges (eigenvalues). However, a
traditional gauge theory of an abelian group, which has no
mathematical distinction between the coupling constant and the
charges, can couple to each field with a different coefficient in
the covariant derivative. So there is problem of charge
quantization in the traditional gauge theory of an abelian group.
\par
Riemannian gauge theory offers a solution
to this problem of traditional gauge theory.
In the definition
(\ref{EqGaugeCovariantDerivativeExpression})
of the covariant derivative,
\(D_\sigma (\phi) = e_a \left( \delta ^a_b
\partial_\sigma
+ \dualrep{e}^a(\partial_\sigma e_b ) \right) \phi^b\),
the relation $(A_\mu)^b_{\ a} = i \,
\dualrep{e}^b \, (\partial_\mu e_a)$
between the gauge field and the basis
vectors has no adjustable parameter
even in the abelian case $(A_\mu)^1_{\ 1} = i \,
\dualrep{e}^1 \, (\partial_\mu e_1)$\@.
This lack of an adjustable parameter is
expected in the covariant derivative of
an $SU(n)$ gauge theory, but it is
surprising in an abelian gauge theory.
\par
Expression (\ref{EqGaugeCovariantDerivativeExpression})
is not our definition.  This expression
follows directly from differential geometry.
The physical motivation to use
differential geometry is to explain the four forces in terms of
symmetry and geometry.
\par
In $U(1)$ gauge theory, the lack
of an adjustable parameter in the
covariant derivative (\ref{EqGaugeCovariantDerivativeExpression})
leads to
$U(1)$ charge quantization. Every
matter-field vector \(\phi^1 \, e_1\)
on the gauge fiber $F$ couples with the
same coefficient. The result is
stronger than charge quantization; it
is charge uniqueness.
The uniqueness of the coupling
in the covariant derivative arises because
the matter fields are defined as
vectors on a flat gauge vector bundle
and because the gauge-covariant
derivative is defined like the covariant
derivative of Riemannian geometry.
Just as geometry fixes the coefficient of
$\Gamma^\lambda_{\mu\nu}$, so too geometry fixes the coefficient of
$A_\mu$\@.
\par
A single adjustable parameter for a
particular vector bundle occurs
as the coefficient of the
field strength term
$q^{-2} {\rm Tr}(F_{\mu \nu} F^{\mu \nu} )$\@.
This coefficient scales the action
of curved gauge fiber.
\par
We have shown that all matter-field
vectors on a gauge vector bundle will
couple in the covariant derivative with
the same coupling constant. This unique
coupling constant does not preclude
positive and negative charges.
Whenever one has two self-conjugate fields
\(\phi^1\) and \(\phi^2\) of the same mass,
one may form the complex field
\(\phi = (1/\sqrt{2})(\phi^1 + i\phi^2)\),
which creates a particle and deletes
its antiparticle; if the particle
has charge \(q\), then the antiparticle
has charge \(- q\)\@.
The particle and antiparticle correspond
to the two solutions
$ D_t \phi = \pm \, i \, \omega \, \phi$\@.
Matter-field vectors $\phi^a e_a$
with opposite charges rotate in time
in opposite directions on the gauge
fiber as shown in figure \ref{FigTwoFieldsOnOneGeometry}
and the attached animation.
Localized matter-fields with opposite
directions of rotation on a background
curved gauge fiber of an
electromagnetic field
accelerate in opposite spatial directions.
\par
One cannot change the charge by defining
$(A_\mu)^a_{\ b}=\frac{1}{q} \,i\,\dualrep{e}^1 \, (\partial_\mu e_1)$\@.
The equations of motion and the wave function evolution
can be solved without ever referencing
the definition (\ref{EqGaugeFieldDefinitionEmbedding}) of $A_\mu$.
Appendix \ref{AppendixSolutionSO2} shows the solution
of the quantum-mechanical states without defining $A_\mu$.
The appendix connects the geometry and the curvature directly to the
evolution of the wave function.  The results of this calculation
are depicted in figures \ref{FigCyclotronSetup} through \ref{FigTwoFieldsOnOneGeometry}.
The energy fixes the rotation rate of the matter vectors in time.  The
canonical momentum fixes the rotation rate of the matter vectors in space.
The only freedom exists in the direction of the rotation (clockwise or counterclockwise) which
leads to positive and negative charges.
\par
If the energy and momentum fix the rate of rotation, then what determines the ``charge" of the
wave packet?  In other words, what determines the radius of the cyclotron orbit of a wave packet moving in a background magnetic field?
The answer is that the trajectory is fixed by the background gauge geometry.
The gauge geometry needs to be generated by some source.
The efficiency with which sources curve gauge fiber is the ``charge" of the theory.
\par
As was explained in section \ref{SecExamples},
 multiple independent $U(1)$
charges can occur in Riemannian gauge theory,
albeit unnaturally.
Using the first method described in section \ref{SecExamples},
one may introduce multiple independent $U(1)$ gauge fibers,
each leading to an independent connection:
$-i\, B_\mu=\dualrep{b}^1(\partial_\mu b_1 )$
and $-i\, C_\mu=\dualrep{c}^1(\partial_\mu c_1
)$.  The basis vectors $b_1$ and $c_1$
are the basis vectors of two \emph{a-priori-}independent,
$1$-complex-dimensional gauge vector
bundles $F^{(B)}$ and $F^{(C)}$.  To
get different charges related to the
same gauge field, we need to constrain
the intrinsic curvature of the two
fibers $F^{(B)}$ and $F^{(C)}$ such
that the connections are related by
$B_\mu / q^{B}= C_\mu / q^C = A_\mu$.
The geometry of two \emph{a-priori-}independent gauge fibers
is then constrained and expressed in terms of
the variable $A_\mu$.  Matter fields on
$F^{B}$ couple with the covariant
derivative $(\partial_\mu - i q^B
A_\mu)$, and matter fields on $F^{C}$
couple with the covariant derivative
$(\partial_\mu - i q^C A_\mu)$.
So by putting different matter fields
on different fibers, we may give them
different charges.
\par
But if all electrons, muons, and taus are
represented by vectors on the same gauge fiber,
then they naturally have the same electric charge.
Thus Riemannian gauge theory shifts the electric charge
quantization problem.  Instead of
wondering why so many $U(1)$ matter
fields couple with the same charge
$q_e$, we ask: Why do the quark
charges differ among themselves
and from the charge of the electron?
Why should matter fields lie on fibers
with curvatures that are related in peculiar ways?
Grandly unified theories provide a
natural way of constraining the
curvature of different fibers.
\par
The $SU(5)$ theory of Georgi and
Glashow~\cite{Georgi:1974sy}
shows how the curvature of
different one-complex-dimensional
fibers may be constrained to produce
different $U(1)$ charges in a theory of grand
unification.
\par
We create a 5-complex-dimensional vector
bundle over space-time spanned by
five orthonormal basis vectors $e_a$.
We restrict $\dualrep e^a (\partial
_\mu e_a) = 0$ to make the group
$SU(5)$ instead of $U(5)$. The
hypercharge gauge field $B_\mu$ is
identified with the gauge field
proportional to the diagonal generator
$\lambda_{24}$
\begin{equation}
-i B_\mu =        -\frac{1}{3} e^1(\partial_\mu e_1)
                  -\frac{1}{3} e^2(\partial_\mu e_2)
                  -\frac{1}{3} e^3(\partial_\mu e_3)
                  +\frac{1}{2} e^4(\partial_\mu e_4)
                  +\frac{1}{2} e^5(\partial_\mu e_5).
\label{EqHyperchargeGenerator}
\end{equation}
%The symmetry breaking process leaves the
%hypercharge generator unbroken.
%The geometry measured by the $B_\mu$ connection is copied
%onto the five $U(1)$ gauge fibers.
The matter-field covariant derivative is still
$D_\mu \phi = e_b \left( \delta^a_{b} \partial_\mu +
e^a(\partial_\mu e_b)\right) \phi^a $,
but only $\frac{1}{3}$ of the term
$e^1(\partial_\mu e_1)$ is identified
as coupling to the hypercharge $B_\mu$\@.
This example shows how $SU(5)$
unification constrains the curvature of
different one-complex-dimensional gauge
fibers so as to give different $U(1)$ charges.
\par
% *************************
% 9 October
% NEW !!!!
% *************************
Ultimately, we must appeal to a method of grand unification to explain different charges.
What then are the new insights into charge quantization?
\par
First, since Riemannian gauge theory
constrains the coupling in the covariant derivative,
the group $U(n)$ may be used in
theories of grand unification without fear of an unconstrained
$U(1)$ subgroup.  Perhaps the actual
unification group is $U(8)_L \times U(8)_R$.
\par
In discussing the equations of general relativity, Einstein called  the geometrical left-hand side ``marble''
and the material right-hand side ``wood.''
Riemannian geometry makes matter into marble -- the matter fields are now geometrical vectors.
In traditional gauge theory, the charge is a free parameter and part of the wood.
Here, there is no charge, only geometry.
\par
Third, we have found the a
level of description where every object is both gauge invariant and
Lorentz invariant.  The gauge fiber and the matter vectors are both gauge invariant
and Lorentz invariant.
\par
Fourth, charge uniqueness is analogous to the
equivalence principle. In Riemannian gauge theory, the
coupling in the covariant derivative is
independent of the field
for the same reason that
in general relativity the gravitational
acceleration is independent of the mass.
The equivalence principle entails
that the manifold is everywhere locally flat,
and therefore that
every small neighborhood can be identified
with a flat tangent space.
Just as the tangent vectors $t_\mu$
describing the tangent space
determine the connection (\ref{EqGRChristoffelDefinition})
of the covariant derivative
by the relation \(\Gamma^\alpha_{\sigma \mu}
= \dualrep{t}^\alpha ( \partial_\sigma  t_\mu)\)
without an adjustable parameter,
so too the basis vectors $e_a$
describing the gauge fiber
determine the connection (\ref{EqGaugeFieldDefinition})
of the gauge-covariant derivative
by the relation \((A_\mu)^a_{\ b} = i
\dualrep{e}^a(\partial_\mu e_b)\)
without an adjustable parameter.
The use of basis vectors to
describe gauge fibers
generalizes the
equivalence principle to gauge theory.
Alternatively,
the observed $U(1)$ charge uniqueness
is the physical principle that motivates
our geometrical description of gauge theory.

\section{Non-compact Gauge Groups}
\label{SecLargerSymmetry}
In section~\ref{SecGeometricalGaugeTheory},
we mentioned that Narasimhan and Ramanan have shown
that the choice of an
orthonormal basis $\langle e_a,e_b
\rangle = \delta_{\bar a b}$ on the gauge
fiber $F|_x$ leads to an $SO(n)$ or $U(n)$ gauge theory.
In eqs.~(\ref{EqPhiAction}--\ref{EqGFAction}),
we wrote the action of a general Riemannian gauge theory
in terms of basis-independent quantities.
A gauge transformation is a change in the
choice of basis vectors used to describe vectors
in the gauge fiber.
Since the action of Riemannian gauge theory
is independent of the choice of basis,
we can choose any linearly independent basis
for the gauge fibers $F={\bf R}^n$ and $F={\bf C}^n$.
\par
When the gauge basis vectors are allowed
to be an arbitrary linearly independent set,
then the symmetry group of the fiber
is $GL(n,{\bf R})$ or $GL(n,{\bf C})$,
and not just $SO(n)$ or $U(n)$\@.
The action of Riemannian gauge theory
and the quantities that follow from it
are invariant under $GL(n,{\bf R})$
or $GL(n,{\bf C})$ gauge transformations.
\par
In traditional gauge theory,
one includes in the action
every term that is renormalizable and gauge invariant
In a traditional gauge
theory~\cite{Cahill:1978ps,Cahill:1979qt}
of the non-compact group
$GL(n,{\bf C})$,
the term
\begin{equation}
S_g = {m^2} \int d^4x\, \sqrt{-g}\,
(D_\mu g)_{a \bar b}\,\,
g^{\bar b c}\, (D^\mu g)_{c\bar d}\, g^{\bar d a}
\label{EqgTerm}
\end{equation}
occurs because it is invariant;
it gives a mass to the gauge bosons associated
with the non-compact generators.
The physical content of this theory changes, however,
when it is interpreted as a
Riemannian gauge theory.
In this case, as we now show,
the term (\ref{EqgTerm}) vanishes;
all the gauge bosons are massless;
and the ones associated
with the non-compact generators
are merely gauge artifacts.
The reason is that
the covariant derivative of the
gauge fiber metric~\cite{Cahill:1978ps,Cahill:1979qt}
\beq
(D_\mu g)_{\bar a b}=\partial_\mu g_{\bar a b}
- i\,(A_\mu)^{\bar c}_{\ \bar a} g_{\bar c b}
+ i \, g_{\bar a c} \, (A_\mu)^c_{\ b}
 \label{EqGaugeCovariantDerivativeMetric}
\eeq
vanishes in Riemannian gauge theory.
If we differentiate the definition
(\ref{EqGaugeFiberMetric}) of the metric
\begin{equation}
  \partial_\mu  g_{\bar a b}
  =
  (\partial_\mu \embedvec{e}_{\bar a}^{\ \bar j})\,
  \embedvec{e}_{b \bar j}
  +
  \embedvec{e}_{\bar a}^{\ \bar j}\,
  (\partial_\mu \embedvec{e}_{b \bar j}),
\end{equation}
and use the metric to raise and lower indices,
\emph{e.g.,}
$\dualrep{e}^c_{\ j} g_{c \bar b} \delta^{j \bar k} =
e_{\bar b}^{\ \bar k}$,
then  with an appropriate complex conjugation,
we find
\begin{equation}
  \partial_\mu  g_{\bar ab}
  =
  (\partial_\mu \embedvec{e}_{\bar a}^{\ \bar j})\,
  \embedvec{e}^{\bar c}_{\ \bar j}\, g_{\bar c b}
  +
  g_{c \bar a}\,\embedvec{e}^{c}_{\ j}\,
  (\partial_\mu \embedvec{e}^{\ j}_{b}) .
\end{equation}
Using the
definition (\ref{EqGaugeFieldDefinitionEmbedding})
of the gauge field,
we see that the covariant derivative of the
metric vanishes, $(D_\mu g)_{\bar a b}=0$,
and so the term (\ref{EqgTerm})
does not contribute to the action
of the Riemannian gauge theory.
\par
The covariant derivative of the
gauge fiber metric vanishes
because the gauge field and
the metric are defined in terms
of basis vectors.
This result is reminiscent of general relativity.
There the covariant derivative of the
space-time metric vanishes
because the connection coefficients and
the metric are defined in terms
of tangent basis vectors.
\par
Freedman, Haagensen, Johnson, Latorre, and
Lam~\cite{Freedman:1993mu,Haagensen:1995sy,Haagensen:1996py}\@.
showed a map from a local $GL(3,{\bf R})$
symmetry group onto an $SO(3)$ gauge theory.
Their work on the hidden spatial-geometry of Yang-Mills theory
is a special case of our Riemannian gauge theory.
They identified directions on
the gauge fiber $F={\bf R}^3$
with spatial directions on the base manifold.
This identification is
possible because the
gauge fiber metric
satisfies $(D_\mu g)_{ a b}=0$.
The resulting curved manifold with a $GL(3,{\bf R})$
local symmetry represents the geometry of the $SO(3)$ gauge fiber.
\par
In summary, the action of
Riemannian gauge theory
eqs.~(\ref{EqPhiAction} -- \ref{EqGFAction})
is basis independent. The choice of
a non-orthonormal basis gives a
larger $GL(n,{\bf R})$ or
$GL(n,{\bf C})$ symmetry without
adding the term (\ref{EqgTerm}) to
the action.

\label{SecQuadraticGRTerm}

\section{Conclusions}
\label{SecConclusions}

Gauge theories traditionally have been
defined by transformation rules for fields
under a symmetry group.
All gauge-invariant, renormalizable
terms are included in the action.
The resulting gauge theories have many
parallels with Riemannian geometry.
In this paper, we have constructed
a gauge theory based upon Riemannian geometry,
which we have called Riemannian gauge theory.
\par
Although drawn from general relativity,
the action of Riemannian gauge theory
in a particular gauge
is the action of traditional gauge theory.
To measure the curvature of the gauge fiber,
it is necessary to use
a term that is quadratic in the
curvature tensor.
A Riemannian gauge theory
with a $U(n)$ or $SO(n)$
gauge symmetry is automatically invariant
under the larger gauge group
$GL(n,{\bf C})$ or $GL(n,{\bf R})$;
no extra terms are needed in the action.
\par
Riemannian gauge theory offers a new insight
to the problem of the quantization
of charge in abelian gauge theories.
The basis vectors $e_a$
describing the gauge fiber
determine the connection (\ref{EqGaugeFieldDefinition})
of the gauge-covariant derivative
by the relation \((A_\mu)^a_{\ b} = i
\dualrep{e}^a(\partial_\mu e_b)\)
without an adjustable parameter.
In general relativity, all particles that
move on a manifold share a common gravitational acceleration.
In Riemannian gauge theory, all fields defined on a common
gauge fiber share the same charge magnitude.
The charge-quantization puzzle is now shifted to asking why
quarks and leptons exist on independent gauge fibers
related in particular ways.
\par
Riemannian gauge theory describes the four forces
in a gauge-invariant, Lorentz-invariant, geometrical
form; enlarges the
gauge group to a non-compact group; connects the equivalence principle with
charge uniqueness; and changes
the nature of the charge-quantization problem in abelian gauge theories.

\acknowledgments

We should like to thank Paul Alsing, Charles Boyer,
Joshua Erlich, Laura Evans, Yang He,
Mark Maneley, Chris Morath,
Jun Song and Stephanie Sposato
for many hours of helpful discussions.
One of us (M.~S.) would like to thank
David Cardimona for his supervision
of this effort.

\bibliographystyle{JHEP}
% \bibliography{gaugegeometry}

\newcommand{\noopsort}[1]{} \newcommand{\printfirst}[2]{#1}
  \newcommand{\singleletter}[1]{#1} \newcommand{\switchargs}[2]{#2#1}
\providecommand{\href}[2]{#2}\begingroup\raggedright\endgroup

\newpage
\appendix

\section{Solution of an $SO(2)$ matter field on a constant magnetic field}
\label{AppendixSolutionSO2}

To solve for the states and the trajectory of a wave packet, we never
have to make use of the definition (\ref{EqGaugeFieldDefinitionEmbedding})
of $A_\mu$ in terms of the basis vectors.
In this section, we find the states in a background magnetic field
without reference to $A_\mu$.  Instead we directly move from the
geometry to the interaction with the matter fields.

We begin with the equation of motion,
\begin{eqnarray}
 \left( (P^l_{\ k} \partial_\mu )
(P^k_{\ j}\partial^\mu )
- \delta^l_{\ j} m_e^2 \right) e_{\ a}^j \phi^a
 =0 \nonumber \\
 \left( (P^k_{\ j} \partial_\mu )
(P^l_{\ k}\partial^\mu )
- \delta^l_{\ j} m_e^2 \right) \conjrep{e}_{\ l}^a \phi_a
 =0 \nonumber
 \label{EqEOMFixedBackground}
\end{eqnarray}
%eq.~(\ref{EqEOMFixedBackground}),
for matter fields in a background gauge
field where $P$ is the projection operator
and $\phi_a=g_{a \bar b}\phi^{\bar b}$\@.
We wish to consider the case of
an $SO(2)$ gauge theory in the presence
of the constant background magnetic field
(\ref{EqBConstRegionSO2}) given in table
\ref{TableSO2FieldStrengths}.

The basis vectors are time independent.  We will solve the
equations of motion for modes that are time harmonic. The $SO(2)$
equation of motion is
  \begin{equation}
   \left( \matrix{ e_{\ 1}^j & e_{\ 2}^j \cr} \right)
\left[ -\left(\matrix{ \vec \nabla &
-{\conjrep{e}}^2_{\ k} \vec \nabla
e^k_{\ 1} \cr
         {\conjrep{e}}^2_{\ k} \vec \nabla
e^k_{\ 1}   & \vec \nabla \cr } \right)
\cdot \left(\matrix{ \vec \nabla &
-{\conjrep{e}}^2_{\ k} \vec \nabla
e^k_{\ 1} \cr
         {\conjrep{e}}^2_{\ k} \vec \nabla
e^k_{\ 1}   & \vec \nabla \cr } \right)
  \left(  \matrix { \phi^1 \cr
            \phi^2 \cr} \right)
  = (\omega^2 - m_e^2 )
  \left( \matrix { \phi^1 \cr
            \phi^2 \cr} \right) \right]
 \end{equation}
where we have used ${\conjrep{e}}^1_{\
k} \vec \nabla e^k_{\
2}=-{\conjrep{e}}^2_{\ k} \vec \nabla
e^k_{\ 1}$.
\par
The coupled differential
equation can be diagonalized into two
decoupled differential equations by
considering the linear combinations
\begin{eqnarray}
 \Phi^{1} = \frac{1}{\sqrt{2}} (\phi^1 - i \phi^2)
 \label{EqCmpxPhiDefFund}\\
 \Phi_{1} =  \frac{1}{\sqrt{2}} (\phi^1 + i \phi^2)
 \label{eqphipmdefinitions} \label{EqCmpxPhiDef}
\end{eqnarray}
which form two equivalent representations of our solution.
We will refer to the complex vector $\Phi^1$ as the primary representation
and or $\Phi_1$ as the dual representation.
\par
Explicitly substituting the basis vectors in eq.~(\ref{EqBConstRegionSO2}) and using
(\ref{EqCmpxPhiDefFund}) and (\ref{EqCmpxPhiDef}) to decouple the
differential equation, we find
% signs fixed on 9 Oct 03 at 1841.
\begin{equation}
(- \nabla^2 + 2 i B_o x \,\partial_y +
B_o^2 x^2 ) \Phi^{1} =
(\omega^2 - m_e^2) \Phi^{1}
\end{equation}
and
\begin{equation}
(- \nabla^2 - 2 i B_o x \,\partial_y +
B_o^2 x^2 ) \Phi_{1} =
(\omega^2 - m_e^2) \Phi_{1}.
\end{equation}
Notice that these two equations happen to be
complex conjugates of each other.

The solutions to this differential equation in the primary representation are
\begin{eqnarray}
  \Phi^{1}_{pos} = \exp( -i(\omega_n t - p_z z - p_y
  y))\Upsilon_n(x-\frac{p_y}{B_o}) \label{eqSolPhiNegMinusOmega} \\
  \Phi^{1}_{neg} = \exp( +i(\omega_n t - p_z z - p_y
  y))\Upsilon_n(x+\frac{p_y}{B_o}) \label{eqSolPhiNegPlusOmega},
\end{eqnarray}
and the solutions in the dual representation are
\begin{eqnarray}
  \Phi_{1\,pos} = \exp( +i(\omega_n t - p_z z - p_y
  y))\Upsilon_n(x-\frac{p_y}{B_o})  \label{eqSolPhiPosPlusOmega} \\
  \Phi_{1\,neg} = \exp( -i(\omega_n t - p_z z - p_y
  y))\Upsilon_n(x+\frac{p_y}{B_o}) \label{eqSolPhiPosMinusOmega}
\end{eqnarray}
where we are using natural units and $\Upsilon_n$ is the $n^{\rm{th}}$
energy eigenstate of the simple harmonic oscillator. The energies are
 \begin{equation}
 \omega_n=+\sqrt{m_e^2 + p_z^2 + 2 | B_o | (n+\frac{1}{2})}.
 \end{equation}

Next, we use eq.~(\ref{EqCmpxPhiDefFund}) and
(\ref{EqCmpxPhiDef}) and rewrite these
solutions in terms of the $SO(2)$ basis vectors $\phi^1$ and $\phi^2$.
For positively charged particles, we have
\begin{eqnarray}
  % Most recently modified 28 Oct 01,
  % Now agrees with A.20 - A.22
   \phi^1_{pos}= {\rm{Re}}( \Phi_{1\,pos})={\rm{Re}}( \Phi^{1}_{pos})= \cos(\omega_n t - p_y y - p_z z) \Upsilon_n(x-\frac{p_y}{B}) \\
   \phi^2_{pos}= {\rm{Im}}( \Phi_{1\,pos})=-{\rm{Im}}( \Phi^{1}_{pos})=\sin(\omega_n t - p_y y - p_z z) \Upsilon_n(x-\frac{p_y}{B}).
\end{eqnarray}
For negatively charged particles, we have
\begin{eqnarray}
   \phi^1_{neg}= {\rm{Re}}( \Phi_{1\,neg})={\rm{Re}}( \Phi^{1}_{neg})= \cos(\omega_n t - p_y y - p_z z) \Upsilon_n(x+\frac{p_y}{B}) \\
   \phi^2_{neg}= {\rm{Im}}( \Phi_{1\,neg})=-{\rm{Im}}( \Phi^{1}_{neg})=-\sin(\omega_n t - p_y y - p_z z) \Upsilon_n(x+\frac{p_y}{B}).
\end{eqnarray}
To generate figures \ref{FigCyclotronSetup} through \ref{FigTwoFieldsOnOneGeometry},
we used the non-relativistic limit of these equations, and
we used a superposition of these states to form the wave packets shown in figure \ref{FigCyclotronSetup}.

In this section, we never made reference
to the definition (\ref{EqGaugeFieldDefinitionEmbedding}) of the connection $A_\mu$.
The interaction was uniquely determined by the curvature.
We observe that all positively charged
solutions will rotate on the plane
spanned by $e^j_{\ 1}$ and $e^j_{\ 2}$
in the opposite direction from the
negatively charged particles.
The rotation in time is determined by the energy of the state;
the rotation as one moves in space is determined by the canonical momentum.
The only freedom one has is to change the direction of rotation which leads to
positive and negative charges.
There is no freedom left to change the magnitude of the charge.

\end{document}